\documentclass[12pt,titlepage]{article}
\usepackage[letterpaper,top=1in,bottom=1in,left=1.25in,right=1.25in]{geometry}
\usepackage{graphicx,cite, url}
\usepackage{color}
\usepackage{threeparttable}
\usepackage{amsmath,amssymb}
\usepackage{makecell}
\usepackage{multirow}
\usepackage{float}
\usepackage{tikz}
\usepackage{algcompatible}
\usepackage{algorithm}
\newcommand*\circled[1]{\tikz[baseline=(char.base)]{
            \node[shape=circle,draw,inner sep=2pt] (char) {#1};}}

\begin{document}
\title{Computational structured illumination for high-content fluorescence and phase microscopy}

\author{Li-Hao Yeh$^{1,2,*}$, Shwetadwip Chowdhury$^{1,2}$ and Laura Waller$^1$\\
\\
\multicolumn{1}{p{\textwidth}}{\centering\emph{\normalsize 1. Department of Electrical Engineering and Computer Sciences, University of California, Berkeley, 94720, USA\\
2. These authors contributed equally to this work\\
$^{*}$  lihao.yeh@berkeley.edu
}}}

\maketitle

\begin{abstract}
High-content biological microscopy targets high-resolution imaging across large fields-of-view (FOVs). Recent works have demonstrated that computational imaging can provide efficient solutions for high-content microscopy. Here, we use speckle structured illumination microscopy (SIM) as a robust and cost-effective solution for high-content fluorescence microscopy with simultaneous high-content quantitative phase (QP). This multi-modal compatibility is essential for studies requiring cross-correlative biological analysis. Our method uses laterally-translated Scotch tape to generate high-resolution speckle illumination patterns across a large FOV. Custom optimization algorithms then jointly reconstruct the sample's super-resolution fluorescent (incoherent) and QP (coherent) distributions, while digitally correcting for system imperfections such as unknown speckle illumination patterns, system aberrations and pattern translations. Beyond previous linear SIM works, we achieve resolution gains of 4$\times$ the objective's diffraction-limited native resolution, resulting in 700 nm fluorescence and 1.2 $\mu$m QP resolution, across a FOV of $2 \times 2.7$ mm$^2$, giving a space-bandwidth product (SBP) of 60 megapixels.
\end{abstract}

\section{Introduction}

The space-bandwidth product (SBP) metric characterizes information content transmitted through an optical system; it can be thought of as the number of resolvable points in an image (\textit{i.e.} the system's field-of-view (FOV) divided by the size of its point spread function (PSF)~\cite{goodman05,Lohmann:96}). Typical microscopes collect images with SBPs of <20 megapixels, a practical limit set by the systems' optical design and camera pixel count. For large-scale biological studies in systems biology and drug discovery, fast high-SBP imaging is desired~\cite{Mccullough2004, MHKim2008, FRDee2009, Pepperkok2006, Yarrow2005, Laketa2007, Trounson2006, Eggert2004}. The traditional solution for increasing SBP is to use an automated translation stage to scan the sample laterally, then stitch together high-content images. However, such capabilities are costly, have long acquisition times and require careful auto-focusing, due to small depth-of-field (DOF) and axial drift of the sample over large scan ranges~\cite{Starkuviene2007}.

Instead of using high-resolution optics and mechanically scanning the FOV, new approaches for high-content imaging use a low-NA objective (with a large FOV) and build up higher resolution by computationally combining a sequence of low-resolution measurements~\cite{Xu2001, Bishara:10, Greenbaum2013, Zheng:2013gq, Tian2014, Tian2015a, Pang2012, Orth2012, Orth2013, Pang2013, Orth2014, Orth2015,Chowdhury2018, kaikai2018}. Such approaches typically illuminate the sample with customized patterns that encode high-resolution sample information into low-resolution features, which can then be measured. These methods reconstruct features smaller than the diffraction limit of the objective, using concepts from synthetic aperture~\cite{Lukosz1967,Schwarz2003,MKim2011} and super-resolution (SR)~\cite{Hell1994, betzig2006imaging, Rust2006, Heintzmann1999, Gustafsson2000, Gustafson2005}. Though the original intent was to maximize resolution, it is important to note that by increasing resolution, SR techniques also increase SBP, and therefore have application in high-content microscopy. Eliminating the requirement for long-distance mechanical scanning means that acquisition is faster and less expensive, while focus requirements are also relaxed by the larger DOF of low-NA objectives.

Existing high-content methods generally use either an incoherent imaging model to reconstruct fluorescence~\cite{Orth2012, Orth2013, Orth2014, Orth2015,Chowdhury2018,Pang2012, Pang2013,kaikai2018}, or a coherent model to reconstruct absorption and quantitative phase (QP)~\cite{Xu2001,Bishara:10,Greenbaum2013,Zheng:2013gq, Tian2014, Tian2015a}. Both have achieved gigapixel-scale SBP (milli-/centi- meter scale FOV with sub-micron resolution). However, none have demonstrated cross-compatibility with both coherent (phase) and incoherent (fluorescence) imaging. Here, we demonstrate multi-modal high-content imaging via a computational imaging framework that allows super-resolution fluorescence and QP. Our method is based on structured illumination microscopy (SIM), which is compatible with both incoherent~\cite{Lukosz1967, Heintzmann1999, Gustafsson2000, dongli2015} and coherent~\cite{vonOlshausen2013, Chowdhury2012, Chowdhury2013, Gao2013, Lee2017, Chowdhury2017RI} sources of contrast~\cite{Chowdhury2017, Chowdhury2017RIFL, Schurmann2017, Shin2018}. 

Though most SIM implementations have focused on super-resolution, some previous works have recognized its suitability for high-content imaging~\cite{Orth2012, Orth2013, Orth2014, Orth2015, Pang2012, Pang2013,Chowdhury2018}. However, these predominantly relied on fluorescence imaging with calibrated illumination patterns, which are difficult to realize in practice because lens-based illumination has finite SBP.  Here, we use random speckle illumination, generated by scattering through Scotch tape, in order to achieve both high-NA \textit{and} large FOV illumination. Our method is related to blind SIM~\cite{Mudry2012}; however, instead of using many random speckle patterns (which restricts resolution gain to $\sim$1.8$\times$), we translate the speckle laterally, enabling resolution gains beyond that of previous methods~\cite{Mudry2012, Ayuk2013, Min2013, Jost2015, Negash2016, Labouesse2016, Yeh2017}  (see Appendix D). Previous works also use high-cost spatial-light-modulators (SLM)~\cite{Forster2014} or galvonemeter/MEMs mirrors~\cite{Dan2013,Lee2017} for precise illumination, as well as expensive objective lenses for aberration correction. We eliminate both of these requirements by performing computational self-calibration, solving for the translation trajectory and the field-dependent aberrations of the system.

Our proposed framework enables three key advantages over existing methods:
\vspace{-5pt}
\begin{itemize}
\setlength{\itemsep}{-2pt}
  \item resolution gains of 4$\times$ the native resolution of the objective (linear SIM is usually restricted to 2$\times$)~\cite{Mudry2012, Ayuk2013, Min2013, Dong2014, Yilmaz2015, Jost2015, Negash2016, Labouesse2016, Yeh2017},
  \item synergistic use of both the fluorescent (incoherent) and quantitative-phase (coherent) signal from the sample to enable multi-modal imaging,
  \item algorithmic self-calibration to significantly relax hardware requirements, enabling low-cost and robust imaging.
\end{itemize}

\begin{figure}[tbh]
\centering
\includegraphics[width=12cm]{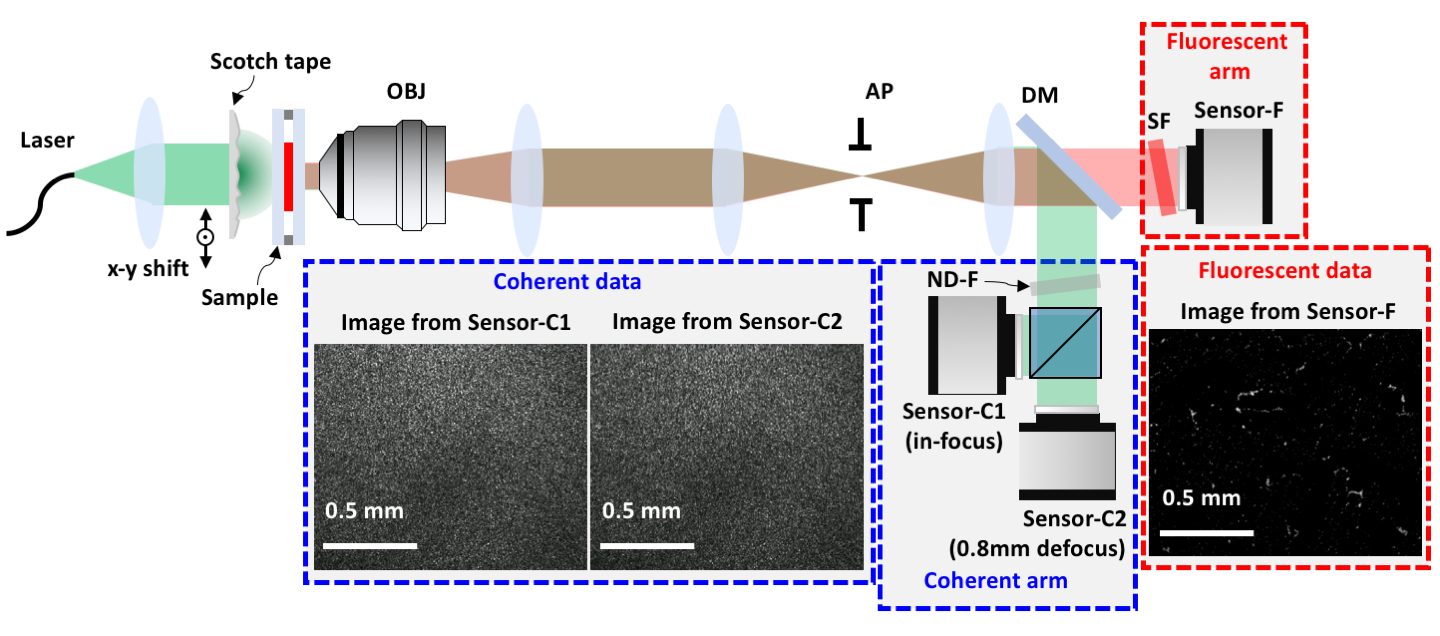}
\caption{Structured illumination microscopy (SIM) with laterally-translated Scotch tape as the patterning element, achieving 4$\times$ resolution gain. Our imaging system has both an incoherent arm, where Sensor-F captures raw fluorescence images (at the emission wavelength, $\lambda_{\mathrm{em}} = 605$ nm) for fluorescence super-resolution, and a coherent arm, where Sensor-C1 and Sensor-C2 capture images with different defocus (at the laser illumination wavelength, $\lambda_{\mathrm{ex}} = 532$ nm) for both super-resolution phase reconstruction and speckle trajectory calibration. OBJ: objective, AP: adjustable iris-aperture, DM: dichroic mirror, SF: spectral filter, ND-F: neutral-density filter.}
\label{fig_setup}
\end{figure}

In our experimental setup, the Scotch tape is placed just before the sample and mounted on a translation stage (Fig.~\ref{fig_setup}). This generates disordered speckles at the sample that are much smaller than the PSF of the imaging optics, encoding SR information. Nonlinear optimization methods are then used to jointly reconstruct multiple calibration quantities: the unknown speckle illumination pattern, the translation trajectory of the pattern, and the field-dependent system aberrations (on a patch-by-patch basis). These are subsequently used to decode the SR information of both fluorescence and phase. Compared to traditional SIM systems that use high-NA objective lenses, our system utilizes a low-NA low-cost lens to ensure large FOV. The Scotch tape generated speckle illumination is not resolution-bound by any imaging lens; this is what allows us to achieve 4$\times$ resolution gains. The result is high-content imaging at sub-micron resolutions across millimeter scale regions. Various previous works have achieved cost-effectiveness, high-content (large SBP), or multiple modalities, but we believe this to be the first to simultaneously encompass all three.

\section{Theory}


SIM generally achieves super-resolution by illuminating the sample with a high spatial-frequency pattern that mixes with the sample's information content to form low-resolution "beat" patterns (\textit{i.e.} moire fringes). Measurements of these "beat" patterns allow elucidation of sample features beyond the diffraction-limited resolution of the imaging system. Maximum achievable resolution in SIM is set by the sum of the numerical apertures (NAs) of the illumination pattern, $\mathrm{NA_{\mathrm{illum}}}$, and the imaging system, $\mathrm{NA_{\mathrm{sys}}}$. Thus, SIM enables a resolution gain factor (over the system's native resolution) of $(\mathrm{NA_{\mathrm{illum}}}+\mathrm{NA_{\mathrm{sys}}})/\mathrm{NA_{\mathrm{sys}}}$ ~\cite{Gustafsson2000}. The minimum resolvable feature size is inversely related to this bound, $d\propto 1/(\mathrm{NA_{\mathrm{illum}}}+\mathrm{NA_{\mathrm{sys}}})$. 

Linear SIM typically maximizes resolution by using either: 1) a high-NA objective in epi-illumination configuration, or 2) two identical high-NA objectives in transmission geometry~\cite{Gustafsson2000,Chowdhury2017}. Both result in a maximum of $2\times$ resolution gain because $\mathrm{NA}_{\mathrm{illum}} = \mathrm{NA}_{\mathrm{sys}}$, which corresponds to an SBP increase by a factor of $4\times$. Given the relatively low native SBP of high-NA imaging lenses, such increases are not sufficient to qualify as high-content imaging. Though nonlinear SIM techniques can enable higher resolution gains~\cite{Gustafson2005}, they require either fluorophore photo-switching or saturation capabilities, which can associate with photobleaching and low SNR, and are not compatible with coherent QP techniques. 

In this work, we aim for $>2\times$  resolution gain; hence, we need the illumination NA to be larger than the detection NA, without using a high-resolution illumination lens (that would restrict the illumination FOV). To achieve this, we use a wide-area high-angle scattering element - layered Scotch tape - on the illumination side of the sample (Fig.~\ref{fig_setup}). Multiple scattering within the tape creates a speckle pattern with finer features than the PSF of the imaging system, \textit{i.e.} $\mathrm{NA}_{\mathrm{illum}} > \mathrm{NA}_{\mathrm{sys}}$. This means that spatial frequencies beyond 2$\times$ the objective's cutoff are mixed into the measurements, which gives a chance to achieve resolution gains greater than two. 

The following sections outline the algorithm that we use to reconstruct large SBP fluorescence and QP images from low-resolution acquisitions of a sample illuminated by a laterally-translating speckle pattern. Unlike conventional SIM reconstruction methods that use analytic linear inversion, our strategy relies instead on joint-variable iterative optimization, where both the sample and illumination speckle (which is unknown) are reconstructed~\cite{Dong2014, Yilmaz2015, kaikai2018}.

\subsection{Super-resolution fluorescence imaging} 

Fluorescence imaging requires an incoherent imaging model. The intensity at the sensor is a low-resolution image of the sample's fluorescent distribution, obeying the system's incoherent resolution limit, $d_{\mathrm{sys}} = \lambda_{\mathrm{em}}/2\mathrm{NA}_{\mathrm{sys}}$, where $\lambda_{\mathrm{em}}$ is the emission wavelength. The speckle pattern generated through the Scotch tape excites the fluorescent sample with features of minimum size $d_{\mathrm{illum}}= \lambda_{\mathrm{ex}}/2\mathrm{NA}_{\mathrm{illum}}$, where $\lambda_{\mathrm{ex}}$ is the excitation wavelength and $\mathrm{NA}_{\mathrm{illum}}$ is set by the scattering angles exiting the Scotch tape. Approximating the excitation and emission wavelengths as similar ($\lambda = \lambda_{\mathrm{ex}} \approx \lambda_{\mathrm{em}}$), the resolution limit of the SIM reconstruction is $d_{\mathrm{SIM}} \approx \lambda/2(\mathrm{NA}_{\mathrm{sys}}+\mathrm{NA}_{\mathrm{illum}})$, with a resolution gain factor of $d_{\mathrm{sys}}/d_{\mathrm{SIM}}$. This factor is mathematically unbounded; however, it will be practically limited by the illumination NA and SNR (see Appendix D). 

\subsubsection{Incoherent forward model for fluorescence imaging} \label{subsec:incohforwmodel}

Plane-wave illumination of the Scotch tape, positioned at the $\ell$-th scan-point, $\mathbf{r}_\ell$, creates a speckle illumination pattern, $p_f(\mathbf{r}-\mathbf{r}_\ell)$, at the plane of the fluorescent sample, $o_f(\mathbf{r})$, where subscript $f$ identifies variables in the fluorescence channel. The fluorescent signal is imaged through the system to give an intensity image at the camera plane:

\begin{eqnarray}
&&I_{f,\ell}(\mathbf{r}) = \left[o_f(\mathbf{r}) \cdot \mathcal{C} \{p_f(\mathbf{r} - \mathbf{r}_\ell)\}\right] \otimes h_f(\mathbf{r}), \;\; \ell = 1, \ldots, N_{\mathrm{img}},
\label{eqn_forward_f}
\end{eqnarray}

\noindent where $\mathbf{r}$ is the 2D spatial coordinates $(x,y)$, $h_f(\mathbf{r})$ is the system PSF, and $N_{\mathrm{img}}$ is the total number of images captured. The subscript $\ell$ describes the acquisition index.

In this formulation, $o_f(\mathbf{r})$, $h_f(\mathbf{r})$, and $I_{f,\ell}(\mathbf{r})$ are 2D $M\times M$-pixel distributions. To accurately model different regions of the pattern translating into the object's $M\times M$ FOV with incrementing $\mathbf{r}_\ell$, we initialize $p_f(\mathbf{r})$ as a $N\times N$ pixel 2D distribution, with $N>M$, and introduce a cropping operator $\mathcal{C}$ to select the $M\times M$ region of the scanning pattern that illuminates the sample. 

\subsubsection{Inverse problem for fluorescence imaging} \label{subsec_inverseproblem}

We next formulate a joint-variable optimization problem to extract SR estimates of the sample, $o_f(\mathbf{r})$, and illumination distributions, $p_f(\mathbf{r})$, from the raw fluorescence measurements, $I_{f,\ell}(\mathbf{r})$, as well as refine the estimate of the system's PSF~\cite{kaikai2018} (aberrations) and speckle translation trajectory, $\mathbf{r}_\ell$. We start with a crude initialization from raw observations of the speckle made using the coherent imaging arm (more details in Sec.~\ref{subsec_registration}). Defining $f_f(o_f, p_f, h_f, \mathbf{r}_1, \ldots, \mathbf{r}_{N_{\mathrm{img}}})$ as a joint-variable cost function that measures the difference between the raw intensity acquisitions and the expected intensities from estimated variables via the forward model, we have:
\begin{eqnarray}
\begin{aligned}
& \underset{o_f, p_f, h_f, \mathbf{r}_1, \ldots, \mathbf{r}_{N_{\mathrm{img}}}}{\text{min}}
& & f_f(o_f, p_f, h_f, \mathbf{r}_1, \ldots, \mathbf{r}_{N_{\mathrm{img}}}) = \sum_{\ell=1}^{N_{\mathrm{img}}} f_{f,\ell}(o_f, p_f, h_f, \mathbf{r}_\ell), \\
& \mathrm{where} & & f_{f,\ell}(o_f, p_f, h_f, \mathbf{r}_\ell) = \sum_\mathbf{r} \left| I_{f,\ell}(\mathbf{r}) - \left[o_f(\mathbf{r}) \cdot \mathcal{C} \{p_f(\mathbf{r} - \mathbf{r}_\ell)\}\right] \otimes h_f(\mathbf{r}) \right|^2.
\end{aligned}
\label{eqn_optimization_f}
\end{eqnarray}

\noindent To solve, a sequential gradient descent~\cite{Yeh2015, Bottou2010} algorithm is used, where the gradient is updated once for each measurement. The sample, speckle pattern, system's PSF and scanning positions are updated by sequentially running through $N_{\mathrm{img}}$ measurements within one iteration. After the sequential update, an extra Nesterov's accelerated update~\cite{Nesterov1983} is included for both the sample and pattern estimate, to speed up convergence. Appendix A contains a detailed derivation of the gradient with respect to the sample, structured pattern, system's PSF and the scanning position based on the linear algebra vectorial notation. The algorithm is described in Appendix B. 


\subsection{Super-resolution quantitative-phase imaging} \label{subsec_theoryintro_ph}

In this section, we present our coherent model for SR quantitative-phase (QP) imaging. A key difference between the QP and fluorescence imaging processes is that the detected intensity at the image plane for coherent imaging is nonlinearly related to the sample's QP~\cite{Chowdhury2012,goodman05}. Thus, solving for a sample's QP from a single intensity measurement is a nonlinear and ill-posed problem. To circumvent this, we use intensity meaurements from two planes, one in-focus and one out-of-focus, to introduce a complex-valued operator that couples QP variations into measurable intensity fluctuations, making the reconstruction well-posed~\cite{N19846,Fienup1982}. The defocused measurements are denoted by a new subscript variable $z$. Figure~\ref{fig_setup} shows our implementation, where two defocused sensors are positioned at $z_0$ and $z_1$ in the coherent imaging arm. 

Generally, the resolution for coherent imaging is roughly half that of its incoherent counterpart~\cite{goodman05} . For our QP reconstruction, the resolution limit is $d_{\mathrm{SIM}}= \lambda_{\mathrm{ex}}/(\mathrm{NA}_{\mathrm{sys}}+\mathrm{NA}_{\mathrm{illum}})$, where the coherent resolution of the native system and the speckle are $d_{\mathrm{sys}} = \lambda_{\mathrm{ex}}/\mathrm{NA}_{\mathrm{sys}}$ and $d_{\mathrm{illum}} = \lambda_{\mathrm{ex}}/\mathrm{NA}_{\mathrm{illum}}$, respectively.

\subsubsection{Coherent forward model for phase imaging}

Assuming an object with 2D complex transmittance function $o_c(\mathbf{r})$ is illuminated by a speckle field, $p_c(\mathbf{r})$, where subscript $c$ refers to the coherent imaging channel, positioned at the $\ell$-th scanning position $\mathbf{r}_\ell$, we can represent the intensity image formed via coherent diffraction as:
\begin{eqnarray}
&&\hspace{-0.3in}I_{c,\ell z}(\mathbf{r}) = \left|\left[ o_c(\mathbf{r}) \cdot \mathcal{C}\left\{p_c(\mathbf{r} - \mathbf{r}_\ell)\right\}\right] \otimes h_{c,z}(\mathbf{r})\right|^2 = |g_{c,\ell z}(\mathbf{r})|^2, \; \ell = 1, \ldots, N_{\mathrm{img}}, \; z = z_0, z_1,
\label{eqn_forward_c}
\end{eqnarray}
where $g_{c,\ell z}(\mathbf{r})$ and $h_{c,z}(\mathbf{r})$ are the complex electric-fields at the imaging plane and the system's coherent PSF at defocus distance $z$, respectively. The comma in the subscript separates the channel index, $c$ or $f$, from the scanning-position and acquisition-number indices, $\ell$ and $z$. $N_{\mathrm{img}}$ here indicates the total number of translations of the Scotch tape. The defocused PSF can be further broken down into $h_{c,z}(\mathbf{r})=h_{c}(\mathbf{r}) \otimes h_{z}(\mathbf{r})$, where $h_{c}(\mathbf{r})$ is the in-focus coherent PSF and $h_{z}(\mathbf{r})$ is the defocus kernel. Similar to Section~\ref{subsec:incohforwmodel}, $I_{c,\ell z}(\mathbf{r})$, $o_c(\mathbf{r})$, and $h_{c,z}(\mathbf{r})$ are 2D distributions with dimensions of $M\times M$ pixels, while $p_c(\mathbf{r})$ is of size $N\times N$ pixels ($N>M$). $\mathcal{C}$ is a cropping operator that selects the sub-region of the pattern that interacts with the sample. The sample's QP distribution is simply the phase of the object's complex transmittance, $\angle o_c(\mathbf{r})$.

\subsubsection{Inverse problem for phase imaging} \label{subsec_inverseproblem_ph}

We now take the raw coherent intensity measurements, $I_{c,\ell z}(\mathbf{r})$, and the registered trajectory, $\mathbf{r}_{\ell z}$, from both of the defocused coherent sensors (more details in Sec.~\ref{subsec_registration}) as input to jointly estimate the sample's SR complex-transmittance function, $o_c(\mathbf{r})$, and illumination complex-field, $p_c(\mathbf{r})$, as well as the aberrations inherent in the system's PSF, $h_c(\mathbf{r})$. The optimization also further refines the scanning trajectory, $\mathbf{r}_{\ell z}$. Based on the forward model, we formulate the joint inverse problem:

\begin{eqnarray}
\begin{aligned}
& \underset{ \substack{ o_c,p_c,h_{c},\mathbf{r}_{1z_0},\mathbf{r}_{1z_1}, \\ \cdots, \mathbf{r}_{N_{\mathrm{img}}z_0},\mathbf{r}_{N_{\mathrm{img}}z_1}}}{\text{minimize}}
& & f_c(o_c,p_c, h_{c}, \mathbf{r}_{1z_0},\mathbf{r}_{1z_1}, \cdots, \mathbf{r}_{N_{\mathrm{img}}z_0},\mathbf{r}_{N_{\mathrm{img}}z_1}) = \sum_{\ell,z} f_{c, \ell z}(o_c, p_c, h_{c},\mathbf{r}_{\ell z}),\\
& \text{where}
& & f_{c, \ell z}(o_c, p_c, h_{c}, \mathbf{r}_{\ell z}) = \sum_{\mathbf{r}}\left| \sqrt{I_{c,\ell z}(\mathbf{r})} - \left|\left[ o_c(\mathbf{r}) \cdot \mathcal{C}\left\{p_c(\mathbf{r} - \mathbf{r}_{\ell z})\right\}\right] \otimes h_{c,z}(\mathbf{r})\right| \right|^2.
\end{aligned}
\label{eqn_optimization_c}
\end{eqnarray}

\noindent Here, we adopt an amplitude-based cost function, $f_c$, which robustly minimizes the distance between the estimated and measured amplitudes in the presence of noise~\cite{gerchberg1971, Fienup1982, Yeh2015}. We optimize the pattern trajectories, $\mathbf{r}_{\ell,z_0}$ and $\mathbf{r}_{\ell,z_1}$, separately for each coherent sensor, in order to account for any residual misalignment or timing-mismatch (see Sec.~\ref{subsec_registration}). As in the fluorescence case, sequential gradient descent~\cite{Yeh2015,Bottou2010} was used to solve this inverse problem.

\subsection{Registration of coherent images} \label{subsec_registration}

Knowledge of the Scotch tape scanning position, $\mathbf{r}_{\ell}$, reduces the complexity of the joint sample and pattern estimation problem and is necessary to achieve SR reconstructions with greater than $2\times$ resolution gain. Because our fluorescent sample is mostly transparent, the main scattering component in the acquired raw data originates from the Scotch tape. Thus, using a sub-pixel registration algorithm \cite{Guizar-Sicairos2008} between successive coherent-camera acquisitions, which are dominated by the scattered speckle signal, is sufficient to initialize the scanning trajectory of the Scotch tape, 

\begin{eqnarray}
&&\mathbf{r}_{\ell z} = \mathcal{R}\left[ I_{c,1 z}(\mathbf{r}), I_{c,\ell z}(\mathbf{r})\right],
\label{eqn_registration}
\end{eqnarray}

\noindent where $\mathcal{R}$ is the registration operator. These initial estimates of $\mathbf{r}_{\ell z}$ are then updated, alongside $o_f(\mathbf{r})$, $o_c(\mathbf{r})$, $p_f(\mathbf{r})$, and $p_c(\mathbf{r})$ using the inverse models described in Sec.~\ref{subsec_inverseproblem} and \ref{subsec_inverseproblem_ph}. In the fluorescence problem described in Sec.~\ref{subsec_inverseproblem}, we only use the trajectory from the in-focus coherent sensor at $z = 0$ for initialization, so we omit the subscript $z$ in $\mathbf{r}_{\ell z}$.

\section{Experimental results}

Figure~\ref{fig_setup} shows our experimental setup. A green laser beam (BeamQ, 532 nm, 200 mW) is collimated through a single lens. The resulting plane wave illuminates the layered Scotch tape (4 layers of 3M 810 Scotch Tape, S-9783), creating a speckle pattern at the sample. The Scotch tape is mounted on a 3-axis piezo-stage (Thorlabs, MAX311D) to enable lateral speckle scanning. The transmitted light from the sample then travels through a 4$f$ system formed by the objective lens (OBJ) and a single lens. In order to control the NA of our detection system (necessary for our verification experiment), an extra 4$f$ system with an adjustable iris-aperture (AP) in the Fourier space is added. Then, the coherent and fluorescent light are optically separated by a dichroic mirror (DM, Thorlabs, DMLP550R), since they have different wavelengths. The fluorescence is further spectrally filtered (SF) before imaging onto Sensor-F (PCO.edge 5.5). The (much brighter) coherent light is ND-filtered and then split by another beam-splitter before falling on the two defocused coherent sensors, Sensor-C1 and Sensor-C2 (FLIR, BFS-U3-200S6M-C). Sensor-C1 is focused on the sample, while Sensor-C2 is defocused by 0.8 mm.

For our initial verification experiments, we use a 40$\times$ objective (Nikon, CFI Achro 40$\times$) with $\mathrm{NA} = 0.65$ as our system's microscope objective (OBJ). Later high-content experimental demonstrations switch to a 4$\times$ objective (Nikon, CFI Plan Achro 4$\times$) with $\mathrm{NA} = 0.1$.

\subsection{Super-resolution verification}
\subsubsection{Fluorescence super-resolution verification}

We start with a proof-of-concept experiment to verify that our method accurately reconstructs a fluorescent sample at resolutions greater than twice the imaging system's diffraction-limit. To do so, we use the higher-resolution objective (40$\times$, NA 0.65) and a tunable Fourier-space iris-aperture (AP) that allows us to artificially reduce the system's NA ($\mathrm{NA}_{\mathrm{sys}}$), and therefore, resolution. With the aperture mostly closed (to $\mathrm{NA}_{\mathrm{sys}} =0.1$), we acquire a low-resolution SIM dataset, which is then used to computationally reconstruct a super-resolved image of the sample with resolution corresponding to an effective NA = 0.4. This reconstruction is then compared to the widefield image of the sample acquired with the aperture open to $\mathrm{NA}_{\mathrm{sys}} =0.4$, for validation.

\begin{figure}[tbh]
\centering
\includegraphics[width=14cm]{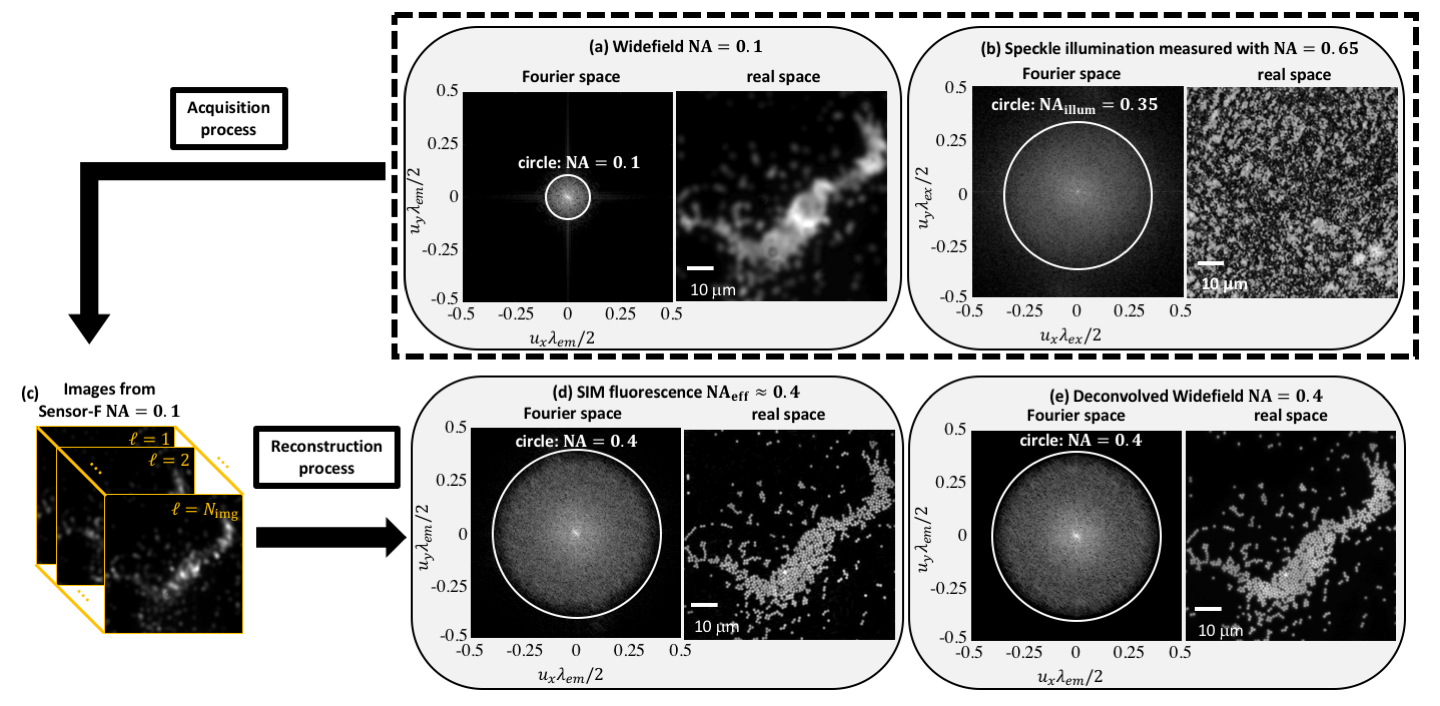}
\caption{Verification of fluorescence super-resolution with 4$\times$ resolution gain. Widefield images, for comparison, were acquired at (a) 0.1 NA and (e) 0.4 NA by adjusting the aperture size. (b) The Scotch tape speckle pattern creates much higher spatial frequencies ($\sim$0.35 NA) than the 0.1 NA detection system can measure. (c) Using the 0.1 NA aperture, we acquire low-resolution fluorescence images for different lateral positions of the Scotch tape. (d) The reconstructed SIM image contains spatial frequencies up to $\sim$0.4 NA and is in agreement with (e) the deconvolved widefield image with the system operating at 0.4 NA.}
\label{fig_verify_fluorescent}
\end{figure}

Results are shown in Fig.~\ref{fig_verify_fluorescent}, comparing our method against widefield fluorescence images at NAs of 0.1 and 0.4, with no Scotch tape in place. The sample is a monolayer of 1 $\mu$m diameter microspheres, with center emission wavelength $\lambda_{\mathrm{em}} = 605$ nm. At 0.1 NA, the expected resolution is $\lambda_{\mathrm{em}}/2$NA $\approx$ 3.0 $\mu$m and the microspheres are completely unresolvable. At 0.4 NA, the expected resolution is $\lambda_{\mathrm{em}}/2$NA $\approx$ 0.76 $\mu$m and the microspheres are well-resolved. With Scotch tape and 0.1 NA, we acquire a set of measurements as we translate the speckle pattern in $267$ nm increments on a $26\times 26$ rectangular grid - $N_{\mathrm{img}}=676$ acquisitions total (details in Sec.~\ref{sec_discussion}). 

Figure~\ref{fig_verify_fluorescent}(d) shows the final SR reconstruction of the fluorescent sample in real space, along with the amplitude of its Fourier spectrum. Individual microspheres can be clearly resolved, and results match well with the 0.4 NA deconvolved widefield image (Fig.~\ref{fig_verify_fluorescent}(e)). Fourier-space analysis confirms our resolution improvement factor to be 4$\times$, which suggests that the Scotch tape produces $\mathrm{NA}_{\mathrm{illum}} \approx 0.3$. To verify, we fully open the aperture and observe that the speckle pattern contains spatial frequencies up to $\mathrm{NA}_{\mathrm{illum}} \approx 0.35$ (Fig.~\ref{fig_verify_fluorescent}(b)).

\subsubsection{Coherent super-resolution verification}

To quantify super-resolution in the coherent imaging channel, we use the low-resolution objective (4$\times$, NA 0.1) to image a USAF1951 resolution chart (Benchmark Technologies). This phase target provides different feature sizes with known phase values, so is a suitable calibration target to quantify both the coherent resolution and the phase sensitivity of our technique. 

\begin{figure}[bth]
\centering
\includegraphics[width=14cm]{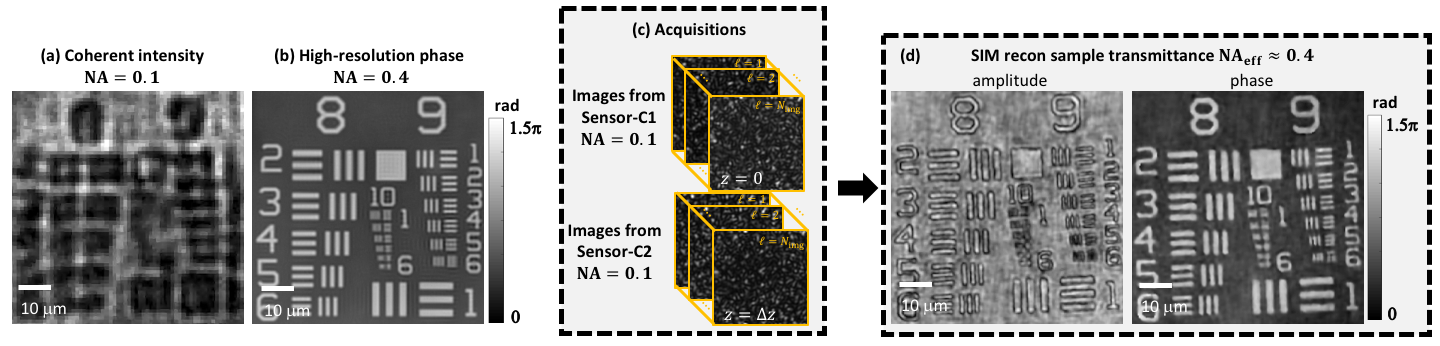}
\caption{Verification of coherent quantitative phase (QP) super-resolution with 4$\times$ resolution gain. \textbf{(a)} Low-resolution intensity image and \textbf{(b)} ``ground truth'' phase at NA=0.4, for comparison. \textbf{(c)} Raw acquisitions of the speckle-illuminated sample intensity from two focus planes, collected with 0.1 NA. \textbf{(d)} Reconstructed SR amplitude and QP, demonstrating 4$\times$ resolution gain.}
\label{fig_verify_phase}
\end{figure}

Results are shown in Fig.~\ref{fig_verify_phase}. The coherent intensity image (Fig.~\ref{fig_verify_phase}(a)) acquired with 0.1 NA (no tape) has low resolution ($\sim 5.32$ $\mu$m), so hardly any features can be resolved . In Fig.~\ref{fig_verify_phase}(b), we show the ``ground truth'' QP distribution at ~0.4 NA, as provided by the manufacturer. 

After inserting the Scotch tape, it was translated in $400$ nm increments on a $36\times 36$ rectangular grid, giving $N_{\mathrm{img}} = 1296$ total acquisitions (details in Sec.~\ref{sec_discussion}) at each of the two defocused coherent sensors (Fig.~\ref{fig_verify_phase}(c)). Figure~\ref{fig_verify_phase}(d,e) shows the SR reconstruction for the amplitude and phase of this sample, resolving features up to group 9 element 5 (1.23 $\mu$m separation). Thus, our coherent reconstruction has a $\sim 4 \times$ resolution gain compared to the brightfield intensity image.

\subsection{High-content multi-modal microscopy}

\begin{figure}[H]
\centering
\includegraphics[width=11cm]{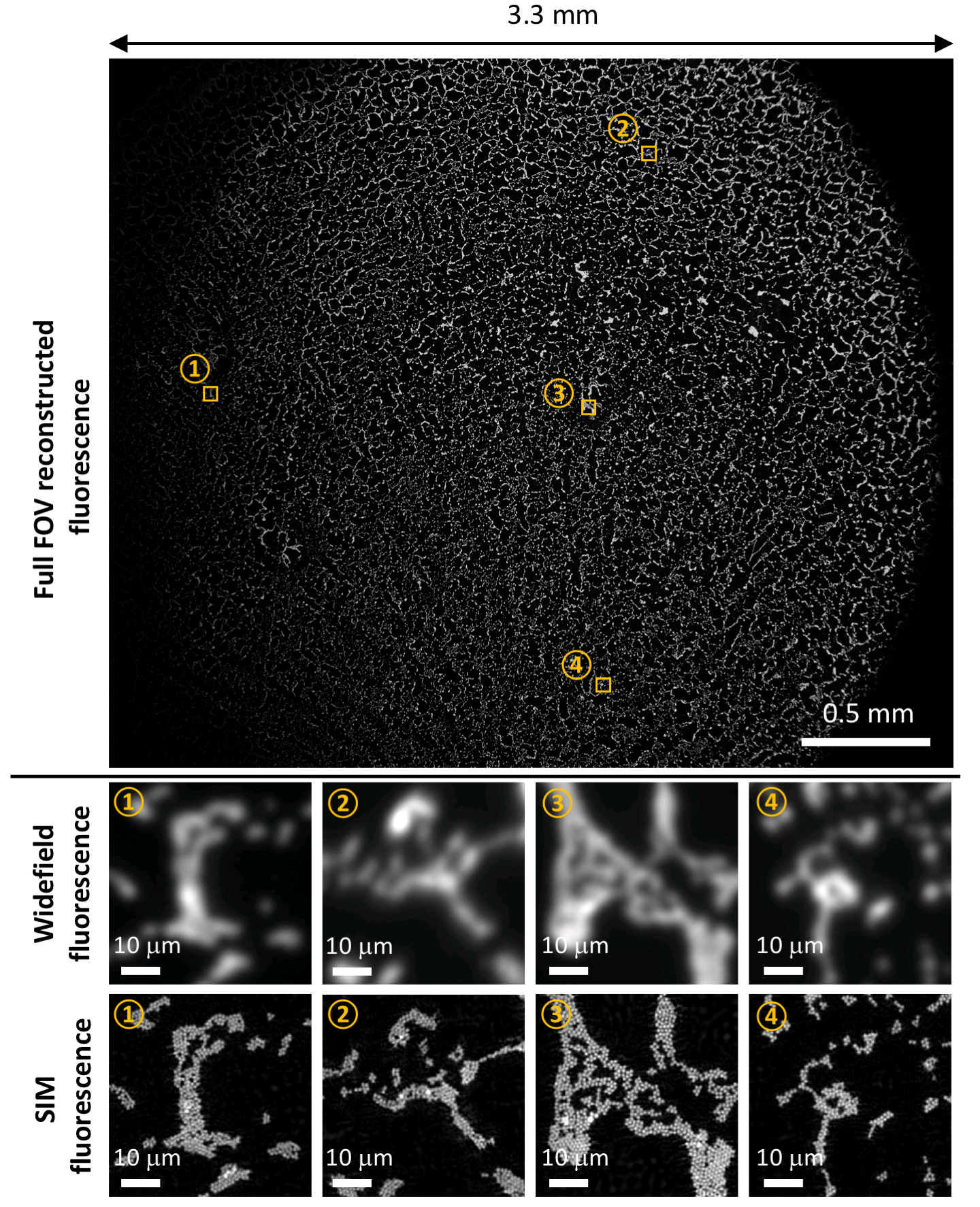}
\caption{Reconstructed super-resolution fluorescence with 4$\times$ resolution gain across the full FOV (See Visualization 1). Four zoom-ins of regions-of-interest (ROIs) are compared to their widefield counterparts.}
\label{fig_large_FOV_FL}
\end{figure}

Of course, artificially reducing resolution in order to validate our method required using a moderate-NA objective, which precluded imaging over the large FOVs allowed by low-NA objectives. In this section, we demonstrate high-content fluorescence imaging with the low-resolution, large FOV objective (4$\times$, NA 0.1) to visualize a 2.7$\times$3.3 mm$^2$ FOV (see Fig.~\ref{fig_large_FOV_FL}(a)). We note that this FOV is more than 100$\times$ larger than that allowed by the 40$\times$ objective used in the verification experiments, so is suitable for large SBP imaging. 

Within the imaged FOV for our 1 $\mu$m diameter microsphere monolayer sample, we zoom in to four regions-of-interest (ROI), labeled \circled{1}, \circled{2}, \circled{3}, and \circled{4}. Widefield fluorescence imaging cannot resolve individual microspheres, as expected. Using our method, however, gives a factor 4$\times$ resolution gain across the whole FOV and enables resolution of individual microspheres. Thus, the SBP of the system, natively $\sim$5.3 mega-pixels of content, was increased to $\sim$85 mega-pixels, a factor of $4^2=16\times$. Though this is still not in the Gigapixel range, this technique is scalable and could reach that range with a higher-SBP objective and sensors.

\begin{figure}[H]
\centering
\includegraphics[width=10cm]{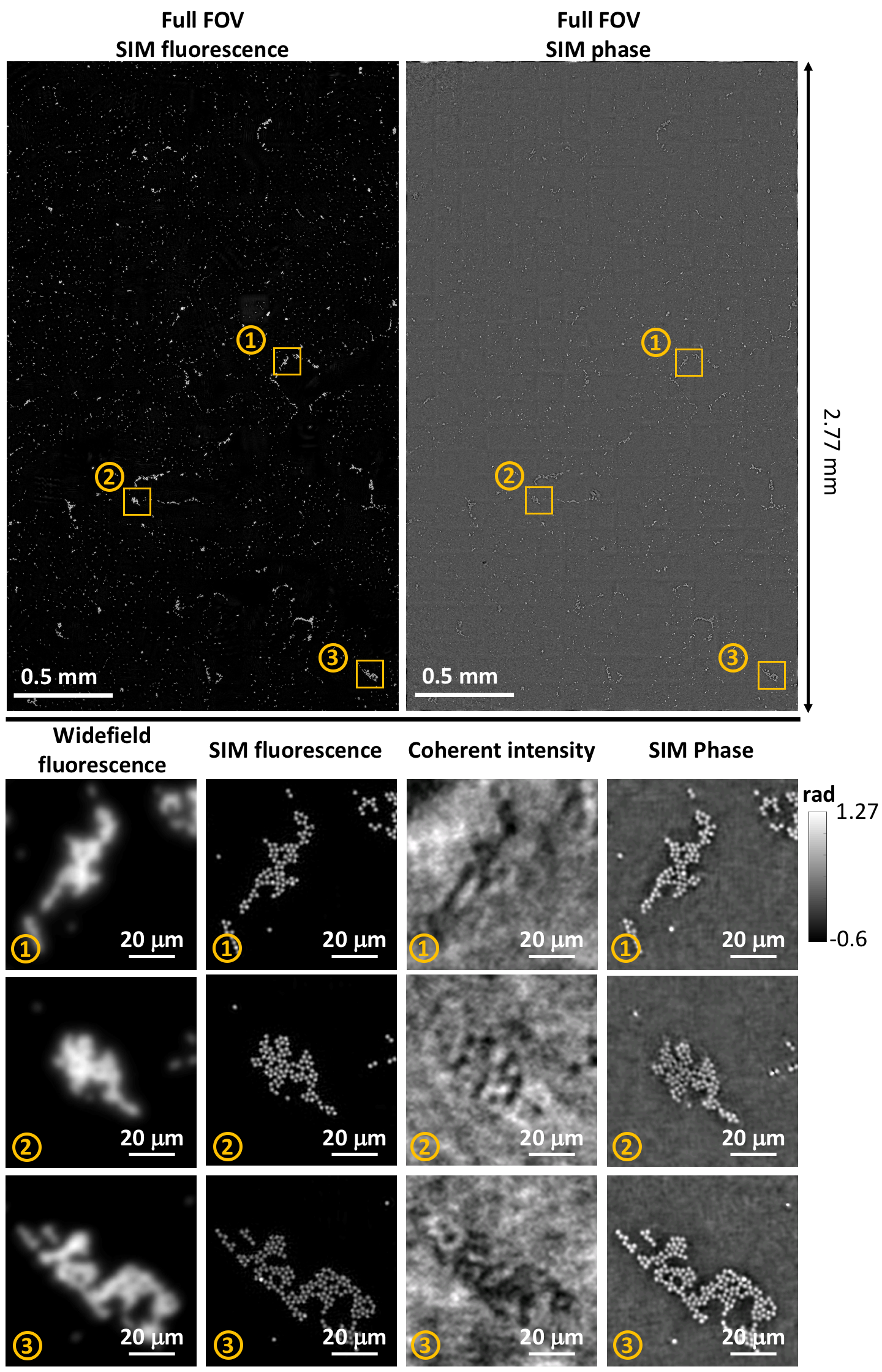}
\caption{Reconstructed multimodal (fluorescence and quantitative phase) high-content imaging (See Visualization 2 and 3). Zoom-ins for three ROIs compare the widefield, super-resolved fluorescence, coherent intensity, and super-resolved phase reconstructions.}
\label{fig_large_FOV_FL_QP}
\end{figure}

We next include the QP imaging channel to demonstrate high-content multimodal imaging, as shown in Fig.~\ref{fig_large_FOV_FL_QP}. The multimodal FOV is smaller (2$\times$2.7 mm$^2$ FOV) than that presented in Fig.~\ref{fig_large_FOV_FL} because our coherent detection sensors have a lower pixel-count than our fluorescence detection sensor. Figure~\ref{fig_large_FOV_FL_QP} includes zoom-ins of three ROIs to visualize the multimodal SR. 

As expected, the widefield fluorescence image and the on-axis coherent intensity image do not allow resolution of individual 2 $\mu$m microspheres, since the theoretical resolution for fluorescence imaging is $\lambda_{\mathrm{em}}/2\mathrm{NA}_{\mathrm{sys}} \approx 3\mu$m and for QP imaging is $\lambda_{\mathrm{ex}}/\mathrm{NA}_{\mathrm{sys}}\approx 5\mu$m. However, our SIM reconstruction with $4\times$ resolution gain enables clear separation of the microspheres in both channels. Our fluorescence and QP reconstructions match well, which is expected since the fluorescent and QP signal originate from identical physical structures in this particular sample. 

The full-FOV reconstructions (Fig.~\ref{fig_large_FOV_FL} and~\ref{fig_large_FOV_FL_QP}) are obtained by dividing the FOV into small patches, reconstructing each patch, then stitching together the high-content images. Patch-wise reconstruction is computationally favorable because of its low-memory requirement, but also allows us to correct field-dependent aberrations. Since we process each patch separately using our self-calibration algorithm, we solve for each patch's PSF independently and correct the local aberrations digitally. The reconstruction takes approximately 15 minutes for each channel on a high-end GPU (NVIDIA, TITAN Xp) for a patch with FOV of $110\times 110$ $\mu$m$^2$. 

\section{Discussion} \label{sec_discussion}

Unlike many existing high-content imaging techniques, one benefit of our method is its easy compatibility for simultaneous QP and fluorescence imaging. This arises from SIM's unique ability to multiplex both coherent and incoherent signals into the system aperture~\cite{Chowdhury2017}. Furthermore, existing high-content fluorescence imaging techniques that use micro-lens arrays~\cite{Orth2012, Orth2013, Orth2014, Orth2015,Pang2012, Pang2013} are resolution-limited by the physical size of the lenslets, which typically have $\mathrm{NA}_{\mathrm{illum}} < 0.3$. Recent work~\cite{Chowdhury2018} has introduced a framework in which gratings with sub-diffraction slits allow sub-micron resolution across large FOVs - however, this work is heavily limited by SNR, due to the primarily opaque grating, as well as tight required axial alignment. Though the Scotch tape used in our proof-of-concept prototype also induced illumination angles within a similar range as micro-lens arrays ($\mathrm{NA}_{\mathrm{illum}} \approx 0.35$), we could in future use a stronger scattering media to achieve $\mathrm{NA}_{\mathrm{illum}} \approx 1.0$, enabling further SR and thus larger SBP. 

The main drawback of our technique is that we use around $\sim1200$ translations of the Scotch tape for each reconstruction, which results in long acquisition times ($\sim$ 180 seconds for shifting, pausing, and capturing) and higher photon requirements. Heuristically, for both fluorescence and QP imaging, we found that a sufficiently large scanning range (larger than $\sim 2$ low-NA diffraction limited spot sizes) and finer scan steps (smaller than the targeted resolution) can reduce distortions in the reconstruction. Tuning such parameters to minimize the number of acquisitions without degrading reconstruction quality is thus an important subject for future endeavors.

\section{Conclusion}
We have presented a large-FOV multimodal SIM fluorescence and QP imaging technique. We use Scotch tape to efficiently generate high-resolution features over a large FOV, which can then be measured with both fluorescent and coherent contrast using a low-NA objective. A computational optimization-based self-calibration algorithm corrected for experimental uncertainties (scanning-position, aberrations, and random speckle pattern) and enabled super-resolution fluorescence and quantitative phase reconstruction with factor $4\times$ resolution gain. 

\newpage
\section*{Appendix A: Gradient derivation}
\subsection*{A.1. Vectorial notation}
\subsubsection*{A.1.1. Fluorescence imaging vectorial model}
In order to solve the multivariate optimization problem in Eq.~\eqref{eqn_optimization_f} and \eqref{eqn_optimization_c} and derive the gradient of the cost function, it is more convenient to consider a linear algebra vectorial notation of the forward models. The fluorescence SIM forward model in Eq.~\eqref{eqn_forward_f} can be alternatively expressed as 
\begin{eqnarray}
&&\mathbf{I}_{f,\ell} = \mathbf{H}_f \mathrm{diag}\left( \mathbf{S}(\mathbf{r}_\ell) \mathbf{p}_f \right) \mathbf{o}_f, 
\label{eqn_forward_f_vec}
\end{eqnarray}

\noindent
where $\mathbf{I}_{f,\ell}$, $\mathbf{H}_f$, $\mathbf{S}(\mathbf{r}_\ell)$, $\mathbf{p}_f$, and $\mathbf{o}_f$ designate the raw fluorescent intensity vector, diffraction-limit low-pass filtering operation, pattern translation/cropping operation, $N^2\times 1$ speckle pattern vector, and $M^2\times 1$ sample's fluorescent distribution vector, respectively. The 2D-array variables described in~\eqref{eqn_forward_f} are all reshaped into column vectors here. $\mathbf{H}_f$ and $\mathbf{S}(\mathbf{r}_\ell)$ can be further broken down into their individual vectorial components:
\begin{eqnarray}
&&\mathbf{H}_f = \mathbf{F}_M^{-1} \mathrm{diag}\left(\tilde{\mathbf{h}}_f \right) \mathbf{F}_M, \nonumber \\
&&\mathbf{S}(\mathbf{r}_\ell) = \mathbf{Q} \mathbf{F}_N^{-1} \mathrm{diag}( \mathbf{e}(\mathbf{r}_\ell)) \mathbf{F}_N,
\label{eqn_forward_f_operator}
\end{eqnarray}
where $\tilde{\mathbf{h}}_f$ is the OTF vector and $\mathbf{e}(\mathbf{r}_\ell)$ is the vectorization of the $\exp(-j2\pi \mathbf{u}\cdot \mathbf{r}_\ell)$ function, where $\mathbf{u}$ is spatial frequency. The notation $\mathrm{diag} (\mathbf{a})$ turns a $n \times 1$ vector, $\mathbf{a}$, into an $n \times n$ diagonal matrix with diagonal entries from the vector entries. $\mathbf{F}_N$ and $\mathbf{F}_M$ denote the $N\times N$-point and $M\times M$-point 2D discrete Fourier transform matrix, respectively, and $\mathbf{Q}$ is the $M^2 \times N^2$ cropping matrix.

With this vectorial notation, the cost function for a single fluorescence measurement is 
\begin{eqnarray}
&&f_{f,\ell}(\mathbf{o}_f, \mathbf{p}_f, \tilde{\mathbf{h}}_f, \mathbf{r}_\ell) = \mathbf{f}_{f,\ell}^T \mathbf{f}_{f,\ell} =  \left \| \mathbf{I}_{f,\ell} - \mathbf{H}_f \mathrm{diag}\left( \mathbf{S}(\mathbf{r}_\ell) \mathbf{p}_f\right) \mathbf{o}_f\right\|_2^2,
\label{eqn_cost_vec_f}
\end{eqnarray}
where $\mathbf{f}_{f,\ell} = \mathbf{I}_{f,\ell} - \mathbf{H}_f \mathrm{diag}\left( \mathbf{S}(\mathbf{r}_\ell) \mathbf{p}_f\right) \mathbf{o}_f$ is the cost vector and $T$ denotes the transpose operation. 

\subsubsection*{A.1.2. Coherent imaging vectorial model}
As with the fluorescence vectorial model, we can rewrite Eq.~\eqref{eqn_forward_c} using vectorial notation: 
\begin{eqnarray}
&&\mathbf{I}_{c,\ell z} = \left| \mathbf{g}_{c,\ell z} \right|^2,
\label{eqn_forward_c_vec}
\end{eqnarray}
where 
\begin{eqnarray}
&&\mathbf{g}_{c,\ell z} = \mathbf{H}_{c,z} \mathrm{diag}(\mathbf{S}(\mathbf{r}_{\ell z}) \mathbf{p}_c) \mathbf{o}_c \nonumber \\
&&\mathbf{H}_{c,z} = \mathbf{F}_M^{-1} \mathrm{diag}(\tilde{\mathbf{h}}_c) \mathrm{diag}(\tilde{\mathbf{h}}_z) \mathbf{F}_M.
\label{eqn_forward_c_vec_def}
\end{eqnarray}
$\mathbf{o}_c$ and $\mathbf{p}_c$ are the $M^2 \times 1$ sample transmittance function vector and $N^2\times 1$ structured field vector, respectively. $\tilde{\mathbf{h}}_c$ and $\tilde{\mathbf{h}}_z$ are the system pupil function and the deliberate defocus pupil function, respectively. With this vectorial notation, we can then express the cost function for a single coherent intensity measurement as 
\begin{eqnarray}
&&f_{c,\ell z}(\mathbf{o}_c, \mathbf{p}_c, \tilde{\mathbf{h}}_c, \mathbf{r}_{\ell z}) = \mathbf{f}_{c,\ell z}^T \mathbf{f}_{c,\ell z} =  \left \| \sqrt{\mathbf{I}_{c,\ell z}} - |\mathbf{g}_{c,\ell z}|\right\|_2^2,
\label{eqn_cost_vec_c}
\end{eqnarray}
where $\mathbf{f}_{c,\ell z} = \sqrt{\mathbf{I}_{c,\ell z}} - |\mathbf{g}_{c,\ell z}|$ is the cost vector for the coherent intensity measurement.

\subsection*{A.2. Gradient derivation}
\subsubsection*{A.2.1. Gradient derivation for fluorescence imaging}
To optimize Eq.~\eqref{eqn_optimization_f} for the variables $\mathbf{o}_f$, $\mathbf{p}_f$, $\tilde{\mathbf{h}}_f$ and $\mathbf{r}_\ell$, we first derive the necessary gradients of the fluorescence cost function. Consider taking the gradient of $f_{f,\ell}$ with respect to $\mathbf{o}_f$, we can represent the $1 \times M^2$ gradient row vector as 
\begin{eqnarray}
&&\frac{\partial f_{f,\ell} }{\partial \mathbf{o}_f} = \left(\frac{\partial f_{f,\ell} }{\partial \mathbf{f}_{f,\ell}}\right) \cdot \left( \frac{\partial \mathbf{f}_{f,\ell}}{\partial \mathbf{o}_f} \right) = \left(2 \mathbf{f}_{f,\ell}^T \right) \cdot \left(- \mathbf{H}_f \mathrm{diag}\left(\mathbf{S}(\mathbf{r}_\ell)\mathbf{p}_f\right) \right).
\label{eqn_grad_o_row_vec}
\end{eqnarray}
Turning the row gradient vector into a $M^2 \times 1$ column vector in order to update the object vector in the right dimension, we the final gradient becomes 
\begin{eqnarray}
&&\nabla_{\mathbf{o}_f} f_{f,\ell} = \left(\frac{\partial f_{f,\ell}}{\partial \mathbf{o}_f} \right)^T = -2 \mathrm{diag} \left(\mathbf{S}(\mathbf{r}_\ell) \mathbf{p}_f\right) \mathbf{H}_f^T \mathbf{f}_{f,\ell}.
\label{eqn_grad_o_column_vec}
\end{eqnarray}

\noindent
To compute the gradient of $\mathbf{p}_f$, we first rewrite the cost vector $\mathbf{f}_{f,\ell}$ as
\begin{eqnarray}
&&\mathbf{f}_{f,\ell} = \mathbf{I}_{f,\ell} - \mathbf{H}_f \mathrm{diag} \left( \mathbf{o}\right) \mathbf{S}(\mathbf{r}_\ell) \mathbf{p}_f.
\label{eqn_cost_vec_p}
\end{eqnarray}
Now, we can write the gradient of the cost function with respect to the pattern vector in row and column vector form as 
\begin{eqnarray}
&&\frac{\partial f_{f,\ell} }{\partial \mathbf{p}_f} = \left(\frac{\partial f_{f,\ell} }{\partial \mathbf{f}_{f,\ell}}\right) \cdot \left(\frac{\partial \mathbf{f}_{f,\ell}}{\partial \mathbf{p}_f} \right) = \left(2 \mathbf{f}_{f,\ell}^T\right) \cdot \left( - \mathbf{H}_f \mathrm{diag}\left(\mathbf{o}_f\right) \mathbf{S}(\mathbf{r}_\ell) \right)\nonumber \\
&&\nabla_{\mathbf{p}_f} f_{f,\ell} = \left(\frac{\partial f_{f,\ell}}{\partial \mathbf{p}_f} \right)^T = -2 \mathbf{S}(\mathbf{r}_\ell)^T \mathrm{diag} \left( \mathbf{o}_f\right) \mathbf{H}_f^T \mathbf{f}_{f,\ell}.
\label{eqn_grad_p_row_column}
\end{eqnarray}

\noindent
Similar to the derivation of the pattern function gradient, it is easier to work with the rewritten form of the cost vector expressed as
\begin{eqnarray}
&&\mathbf{f}_{f,\ell} = \mathbf{I}_{f, \ell} - \mathbf{F}_M^{-1} \mathrm{diag}\left(\mathbf{F}_M \mathrm{diag}\left(\mathbf{S}(\mathbf{r}_\ell)\mathbf{p}_f \mathbf{o}_f\right)\right) \tilde{\mathbf{h}}_f.
\label{eqn_cost_vec_h}
\end{eqnarray}
The gradient of the cost function with respect to the OTF vector in the row and column vector form are expressed, respectively, as
\begin{eqnarray}
&&\frac{\partial f_{f,\ell}}{\partial \tilde{\mathbf{h}}_f} = \left(\frac{\partial f_{f,\ell}}{\partial \mathbf{f}_{f,\ell}}\right) \cdot \left(\frac{\partial \mathbf{f}_{f,\ell}}{\partial \tilde{\mathbf{h}}_f}\right) = \left(2 \mathbf{f}^T_{f,\ell}\right) \cdot \left(-\mathbf{F}_M^{-1} \mathrm{diag}\left(\mathbf{F}_M \mathrm{diag}\left(\mathbf{S}(\mathbf{r}_\ell)\mathbf{p}_f \mathbf{o}_f\right)\right) \right) \nonumber \\
&&\nabla_{\tilde{\mathbf{h}}_f}f_{f,\ell} = \left(\frac{\partial f_{f,\ell}}{\partial \tilde{\mathbf{h}}_f}\right)^\dagger = -2 \mathrm{diag}\left( \overline{\mathbf{F}_M \mathrm{diag}\left(\mathbf{S}(\mathbf{r}_\ell)\mathbf{p}_f \mathbf{o}_f\right)}\right) \mathbf{F}_M \mathbf{f}_{f,\ell},
\label{eqn_grad_h_row_column}
\end{eqnarray}
where $\overline{\mathbf{a}}$ denotes entry-wise complex conjugate operation on any general vector $\mathbf{a}$. One difference between this gradient and the previous one is that the variable to solve, $\tilde{\mathbf{h}}_f$, is now a complex vector. When turning the gradient row vector of a complex vector into a column vector, we have to take a Hermitian operation, $\dagger$, on the row vector following the conventions in~\cite{Ken2009}. We will have more examples of complex variables in the coherent model gradient derivation.

For taking the gradient of the scanning position, we again rewrite the cost vector $\mathbf{f}_{f,\ell}$: 
\begin{eqnarray}
&&\mathbf{f}_{f,\ell} = \mathbf{I}_\ell - \mathbf{H}_f \mathrm{diag}\left( \mathbf{o}_f\right) \mathbf{Q} \mathbf{F}_N^{-1} \mathrm{diag} \left( \mathbf{F}_N \mathbf{p}_f\right) \mathbf{e}(\mathbf{r}_\ell).
\label{eqn_cost_vec_r}
\end{eqnarray}
We can then write the gradient of the cost function with respect to the scanning position as 
\begin{eqnarray}
&&\hspace{-0.8in}\frac{\partial f_{f,\ell} }{\partial q_\ell} = \left(\frac{\partial f_{f,\ell} }{\partial \mathbf{f}_{f,\ell}} \right) \cdot \left( \frac{\partial \mathbf{f}_{f,\ell}}{\partial \mathbf{e} (\mathbf{r}_\ell)} \right) \cdot \left( \frac{\partial \mathbf{e}(\mathbf{r}_\ell)}{\partial q_\ell} \right) \nonumber \\
&&= \left(2 \mathbf{f}_{f,\ell}^T \right) \cdot \left( - \mathbf{H}_f \mathrm{diag}\left(\mathbf{o}_f\right) \mathbf{Q}\mathbf{F}_N^{-1} \mathrm{diag}\left( \mathbf{F}_N\mathbf{p}_f\right) \right) \cdot \left( \mathrm{diag}\left(-j2\pi \mathbf{u}_q\right)  \mathbf{e}(\mathbf{r}_\ell) \right),
\label{eqn_grad_xy_vec}
\end{eqnarray}
where $q$ is either the $x$ or $y$ spatial coordinate component of $\mathbf{r}_\ell$. $\mathbf{u}_q$ is the $N^2 \times 1$ vectorial notation of the spatial frequency function in the $q$ direction. 

To numerically evaluate these gradients, we represent them in the functional form as:
\begin{eqnarray}
&&\hspace{-0.6in}\nabla_{o_f} f_{f,\ell} (o_f, p_f, h_f, \mathbf{r}_\ell) = -2 p_f(\mathbf{r}-\mathbf{r}_\ell) \cdot \left[ h_f^*(-\mathbf{r}) \otimes \left(I_{f,\ell}(\mathbf{r}) - \left[o_f(\mathbf{r}) \cdot \mathcal{C} \{p_f(\mathbf{r} - \mathbf{r}_\ell)\}\right] \otimes h_f(\mathbf{r})\right)\right], \nonumber \\
&&\hspace{-0.6in}\nabla_{p_f} f_{f,\ell} (o_f, p_f, h_f, \mathbf{r}_\ell) = -2 \delta(\mathbf{r}+\mathbf{r}_\ell) \otimes \mathcal{P}\left\{o_f(\mathbf{r}) \cdot \left[ h_f^*(-\mathbf{r}) \otimes \left(I_{f,\ell}(\mathbf{r}) - \left[o_f(\mathbf{r}) \cdot \mathcal{C} \{p_f(\mathbf{r} - \mathbf{r}_\ell)\}\right] \otimes h_f(\mathbf{r})\right)\right] \right\}, \nonumber \\
&&\hspace{-0.6in}\nabla_{\tilde{h}_f} f_{f,\ell}(o_f, p_f, h_f, \mathbf{r}_\ell) = -2 \left(\mathcal{F}\left\{ o_f(\mathbf{r})\cdot \mathcal{C}\left\{p_f(\mathbf{r}-\mathbf{r}_\ell)\right\}\right\}\right)^* \cdot \mathcal{F}\left\{ I_{f,\ell}(\mathbf{r}) - \left[o_f(\mathbf{r})\cdot \mathcal{C}\left\{p_f(\mathbf{r}-\mathbf{r}_\ell)\right\}\right] \otimes h_f(\mathbf{r})\right\}, \nonumber \\
&&\hspace{-0.6in}\nabla_{q_{\ell}} f_{f,\ell} (o_f, p_f, h_f, \mathbf{r}_\ell) = -2 \left\{\sum_{\mathbf{r}} \left(I_{f,\ell}(\mathbf{r}) - \left[o_f(\mathbf{r}) \cdot \mathcal{C} \{p_f(\mathbf{r} - \mathbf{r}_\ell)\}\right] \otimes h_f(\mathbf{r})\right) \cdot \right. \nonumber \\ 
&&\hspace{1.6 in}\left. h_f(\mathbf{r}) \otimes \left[o_f(\mathbf{r}) \cdot \mathcal{C}\left\{ \frac{\partial p_f(\mathbf{r} - \mathbf{r}_\ell)}{\partial q_\ell}\right\} \right] \right\}, 
\label{eqn_grad_opr}
\end{eqnarray}
where $a^*$ stands for complex conjugate of any general function, $a$, $\mathcal{F}$ is the Fourier transform operator, and $\mathcal{P}$ is a zero-padding operator that pads an $M\times M$ image to size $N\times N$ pixels. In this form, $I_{f,\ell}(\mathbf{r})$, $o_f(\mathbf{r})$, and $h_f(\mathbf{r})$ are 2D $M\times M$ images, while $p_f(\mathbf{r})$ is a $N\times N$ image. The gradients for the sample and the structured pattern are of the same size as $o_f(\mathbf{r})$ and $p_f(\mathbf{r})$, respectively. Ideally, the gradient of the the scanning position in each direction is a real number. However, due to imperfect implementation of the discrete differentiation in each direction, the gradient will have small imaginary value that will be dropped in the update of the scanning position. 

\subsubsection*{A.2.2. Gradient derivation for coherent imaging}
For the coherent imaging case, we will derive the gradients of the cost function in Eq.~\eqref{eqn_cost_vec_c} with respect to the sample transmittance function $\mathbf{o}_c$, speckle field $\mathbf{p}_c$, pupil function $\tilde{\mathbf{h}}_c$, and the scanning position $\mathbf{r}_{\ell z}$. First, we take the gradient of $f_{c,\ell z}$ with respect to $\mathbf{o}_c$, we then have the gradient in the row and column vector forms as
\begin{eqnarray}
&&\frac{\partial f_{c,\ell z}}{\partial \mathbf{o}_c} = \left( \frac{\partial f_{c,\ell z}}{\partial \mathbf{f}_{c,\ell z}} \right) \cdot \left( \frac{\partial \mathbf{f}_{c,\ell z}}{\partial \mathbf{g}_{c,\ell z}} \right) \cdot \left( \frac{\partial \mathbf{g}_{c,\ell z}}{\partial \mathbf{o}_c} \right) \nonumber \\
&&\hspace{0.5 in} = \left( 2\mathbf{f}_{c,\ell z}^T \right) \cdot \left( -\frac{1}{2} \mathrm{diag}\left( \frac{\overline{\mathbf{g}_{c,\ell z}}}{|\mathbf{g}_{c,\ell z}|} \right) \right)\cdot \left( \mathbf{H}_{c,z}\mathrm{diag}\left(\mathbf{S}(\mathbf{r}_{\ell z})\right) \mathbf{p}_c \right) \nonumber \\
&&\nabla_{\mathbf{o}_c} f_{c,\ell z} = \left( \frac{\partial f_{c,\ell z}}{\partial \mathbf{o}_c} \right)^\dagger = -\mathrm{diag}(\overline{\mathbf{S}(\mathbf{r}_{\ell z}) \mathbf{p}_c}) \mathbf{H}_{c,z}^\dagger \mathrm{diag}\left(\frac{\mathbf{g}_{c,\ell z}}{\left|\mathbf{g}_{c,\ell z} \right|}\right) \mathbf{f}_{c,\ell z},
\label{eqn_grad_oc_row_column}
\end{eqnarray}
where the $\frac{\mathbf{g}_{c,\ell z}}{\left|\mathbf{g}_{c,\ell z} \right|}$ operation denotes entry-wise division between the two vectors, $\mathbf{g}_{c,\ell z}$ and $|\mathbf{g}_{c,\ell z}|$. In addition, the detailed calculation of $\frac{\partial \mathbf{f}_{c,\ell z}}{\partial \mathbf{g}_{c,\ell z}}$ can be found in the Appendix of~\cite{Yeh2015}.

\noindent
Next, we take the gradient with respect to the pattern field vector, $\mathbf{p}_c$, and write down the corresponding row and column vectors as 
\begin{eqnarray}
&&\frac{\partial f_{c,\ell z}}{\partial \mathbf{p}_c} = \left( \frac{\partial f_{c,\ell z}}{\partial \mathbf{f}_{c,\ell z}} \right) \cdot \left( \frac{\partial \mathbf{f}_{c,\ell z}}{\partial \mathbf{g}_{c,\ell z}} \right) \cdot \left( \frac{\partial \mathbf{g}_{c,\ell z}}{\partial \mathbf{p}_c} \right) \nonumber \\
&&\hspace{0.5 in} = \left( 2\mathbf{f}_{c,\ell z}^T \right) \cdot \left( -\frac{1}{2} \mathrm{diag}\left( \frac{\overline{\mathbf{g}_{c,\ell z}}}{|\mathbf{g}_{c,\ell z}|} \right) \right)\cdot \left( \mathbf{H}_{c,z}\mathrm{diag}\left(\mathbf{o}_c \right) \mathbf{S}(\mathbf{r}_{\ell z})\right) \nonumber \\
&&\nabla_{\mathbf{p}_c} f_{c,\ell z} = \left( \frac{\partial f_{c,\ell z}}{\partial \mathbf{p}_c} \right)^\dagger = -\mathbf{S}(\mathbf{r}_{\ell z})^\dagger \mathrm{diag}\left(\overline{\mathbf{o}_c}\right) \mathbf{H}_{c,z}^\dagger \mathrm{diag}\left(\frac{\mathbf{g}_{c,\ell z}}{\left|\mathbf{g}_{c,\ell z} \right|}\right) \mathbf{f}_{c,\ell z}.
\label{eqn_grad_pc_row_column}
\end{eqnarray}
In order to calculate $\frac{\partial \mathbf{g}_{c,\ell z}}{\partial \mathbf{p}_c}$, we need to reorder the dot multiplication of $\mathbf{o}_c$ and $\mathbf{S}(\mathbf{r}_{\ell z}) \mathbf{p}_c$ as we did in deriving the gradient of the pattern for fluorescence imaging.

\noindent
In order to do aberration correction, we will need to estimate the system pupil function, $\tilde{\mathbf{h}}_c$. The gradient with respect to the pupil function can be derived as,
\begin{eqnarray}
&&\hspace{-0.8 in}\frac{\partial f_{c,\ell z}}{\partial \tilde{\mathbf{h}}_c} = \left( \frac{\partial f_{c,\ell z}}{\partial \mathbf{f}_{c,\ell z}} \right) \cdot \left( \frac{\partial \mathbf{f}_{c,\ell z}}{\partial \mathbf{g}_{c,\ell z}} \right) \cdot \left( \frac{\partial \mathbf{g}_{c,\ell z}}{\partial \tilde{\mathbf{h}}_c} \right) \nonumber \\
&&\hspace{-0.3 in} = \left( 2\mathbf{f}_{c,\ell z}^T \right) \cdot \left( -\frac{1}{2} \mathrm{diag}\left( \frac{\overline{\mathbf{g}_{c,\ell z}}}{|\mathbf{g}_{c,\ell z}|} \right) \right)\cdot \left( \mathbf{F}_M^{-1} \mathrm{diag}\left[\mathbf{F}_M \mathrm{diag}\left(\mathbf{S}(\mathbf{r}_{\ell z}) \mathbf{p}_c \right) \mathbf{o}_c \right] \mathrm{diag}(\tilde{\mathbf{h}}_z)\right) \nonumber \\
&&\hspace{-0.8 in}\nabla_{\tilde{\mathbf{h}}_c}f_{c,\ell z} = \left(\frac{\partial f_{c,\ell z}}{\partial \tilde{\mathbf{h}}_c}\right)^\dagger = -\mathrm{diag}(\overline{\tilde{\mathbf{h}}_z}) \mathrm{diag}\left[ \overline{\mathbf{F}_M\mathrm{diag}\left(\mathbf{S}(\mathbf{r}_{\ell z}) \mathbf{p}_c\right)\mathbf{o}_c}\right]\mathbf{F}_M \mathrm{diag}\left(\frac{\mathbf{g}_{c,\ell z}}{|\mathbf{g}_{c,\ell z}|}\right) \mathbf{f}_{c,\ell z}.
\label{eqn_grad_hc_row_column}
\end{eqnarray}

In the end, the gradient of the scanning position for refinement can be derived as
\begin{eqnarray}
&&\hspace{-0.8 in}\frac{\partial f_{c,\ell z}}{\partial q_{\ell z}} = \left( \frac{\partial f_{c,\ell z}}{\partial \mathbf{f}_{c,\ell z}} \right) \cdot \left[ \left( \frac{\partial \mathbf{f}_{c,\ell z}}{\partial \mathbf{g}_{c,\ell z}} \right) \cdot \left( \frac{\partial \mathbf{g}_{c,\ell z}}{\partial \mathbf{e}(\mathbf{r}_{\ell z})} \right) \cdot \left(\frac{\mathbf{e}(\mathbf{r}_{\ell z})}{\partial q_\ell}\right) + \left( \frac{\partial \mathbf{f}_{c,\ell z}}{\partial \overline{\mathbf{g}_{c,\ell z}}} \right) \cdot \left( \frac{\partial \overline{\mathbf{g}_{c,\ell z}}}{\partial \overline{\mathbf{e}(\mathbf{r}_{\ell z})}} \right) \cdot \left(\frac{\overline{\mathbf{e}(\mathbf{r}_{\ell z})}}{\partial q_\ell}\right) \right]\nonumber \\
&& \hspace{-0.3in}= 2 \left( \frac{\partial f_{c,\ell z}}{\partial \mathbf{f}_{c,\ell z}} \right) \cdot \mathrm{Re} \left\{ \left( \frac{\partial \mathbf{f}_{c,\ell z}}{\partial \mathbf{g}_{c,\ell z}} \right) \cdot \left( \frac{\partial \mathbf{g}_{c,\ell z}}{\partial \mathbf{e}(\mathbf{r}_{\ell z})} \right) \cdot \left(\frac{\mathbf{e}(\mathbf{r}_{\ell z})}{\partial q_\ell}\right) \right\} \nonumber \\
&&\hspace{-0.3in} = 2 \left( 2\mathbf{f}_{c,\ell z}^T \right) \cdot \mathrm{Re}\left\{\left( -\frac{1}{2} \mathrm{diag}\left( \frac{\overline{\mathbf{g}_{c,\ell z}}}{|\mathbf{g}_{c,\ell z}|} \right) \right)\cdot \left( \mathbf{H}_{c,z}\mathrm{diag}\left(\mathbf{o}_c  \right) \mathbf{Q}\mathbf{F}_N^{-1} \mathrm{diag}\left( \mathbf{F}_N\mathbf{p}_c\right) \right)  \cdot \left( \mathrm{diag}\left(-j2\pi \mathbf{u}_q\right)  \mathbf{e}(\mathbf{r}_{\ell z}) \right)\right\} \nonumber \\
&&\hspace{-0.3in} =  -2\mathrm{Re}\left\{ \mathbf{f}_{c,\ell z}^T \mathrm{diag}\left(\frac{\overline{\mathbf{g}_{c,\ell z}}}{\left|\mathbf{g}_{c,\ell z} \right|}\right) \mathbf{H}_{c,z} \mathrm{diag}(\mathbf{o}_c)\mathbf{Q}\mathbf{F}_N^{-1} \mathrm{diag}(\mathbf{F}_N\mathbf{p}_c) \mathrm{diag}(-j2\pi\mathbf{u}_q) \mathbf{e}(\mathbf{r}_{\ell z}) \right\},
\label{eqn_grad_xy_coh}
\end{eqnarray}
where $q$ is either the $x$ or $y$ spatial coordinate component of $\mathbf{r}_{\ell z}$.

In order to numerically evaluate these gradients, we represent them, as we did for the gradients of the fluorescence model, into functional forms:
\begin{eqnarray}
&&\nabla_{o_c} f_{c,\ell z}(o_c,p_c,h_c,\mathbf{r}_{\ell z}) = -p_c^*(\mathbf{r}-\mathbf{r}_{\ell z}) \cdot \left[ h_{c,z}^*(-\mathbf{r}) \otimes \left( \left( \frac{\sqrt{I_{c,\ell z}(\mathbf{r})} }{|g_{c,\ell z}(\mathbf{r})|} - 1\right) \cdot g_{c,\ell z}(\mathbf{r}) \right)\right] \nonumber \\
&&\nabla_{p_c} f_{c,\ell z}(o_c,p_c,h_c,\mathbf{r}_{\ell z}) = -\delta(\mathbf{r}+\mathbf{r}_{\ell z}) \otimes \mathcal{P}\left\{ o_c^*(\mathbf{r}) \cdot \left[ h_{c,z}^*(-\mathbf{r})\otimes \left( \left( \frac{\sqrt{I_{c,\ell z}(\mathbf{r})} }{|g_{c,\ell z}(\mathbf{r})|} - 1\right) \cdot g_{c,\ell z}(\mathbf{r}) \right) \right]\right\} \nonumber \\
&&\nabla_{\tilde{h}_c} f_{c,\ell z}(o_c,p_c,h_c,\mathbf{r}_{\ell z}) = -\tilde{h}_z^*(\mathbf{u}) \cdot \mathcal{F}\left\{p_c(\mathbf{r}-\mathbf{r}_{\ell z}) \cdot o_c(\mathbf{r}))\right\}^* \mathcal{F}\left\{ \left( \frac{\sqrt{I_{c,\ell z}(\mathbf{r})} }{|g_{c,\ell z}(\mathbf{r})|} - 1\right) \cdot g_{c,\ell z}(\mathbf{r})\right\} \nonumber \\
&&\nabla_{q_{\ell z}} f_{c,\ell z}(o_c,p_c,h_c,\mathbf{r}_{\ell z}) = -2\mathrm{Re}\left\{ \sum_{\mathbf{r}} \left[ \left( \frac{\sqrt{I_{c,\ell z}(\mathbf{r})} }{|g_{c,\ell z}(\mathbf{r})|} - 1\right) \cdot g^*_{c,\ell z}(\mathbf{r}) \right] \cdot \right. \nonumber \\
&&\left. \hspace{2.5 in} \left[ h_{c,z}(\mathbf{r}) \otimes \left( o_c(\mathbf{r}) \cdot \mathcal{C}\left\{ \frac{\partial p_c(\mathbf{r} - \mathbf{r}_{\ell z})}{\partial q_{\ell z}}\right\} \right)\right] \right\}.
\label{eqn_grad_opr_coh}
\end{eqnarray}

\section*{Appendix B: Reconstruction algorithm}

With the derivation of the gradients in Appendix A, we summarize here the reconstruction algorithm for fluorescence imaging and coherent imaging. 

\subsection*{B.1. Algorithm for fluorescence imaging}
First, we initialize the sample, $o_f(\mathbf{r})$, with the mean image of all the structure illuminated images, $I_{f,\ell}(\mathbf{r})$, which is approximately a widefield diffraction-limited image. As for the structured pattern, $p_f(\mathbf{r})$, we initialize it with a all-one image. The initial OTF, $\tilde{h}_f(\mathbf{u})$, is set as a non-aberrated incoherent OTF. Initial scanning positions are from the registration of the in-focus coherent speckle images, $I_{c,\ell z}(\mathbf{r})$ ($z=0$). 

In the algorithm, $K_f$ is the total number of iterations ($K_f = 100$ is generally enough for convergence). At every iteration, we sequentially update the sample, structured pattern, system's OTF and the scanning position using each single frame from $\ell = 1$ to $\ell = N_{\mathrm{img}}$. A Nesterov acceleration step is applied on the sample and the structured pattern at the end of each iteration. The detailed algorithm is summarized in Algorithm~\ref{algo_fluor}.

\begin{algorithm} [H]                    
\caption{Fluorescence imaging reconstruction}
\label{algo_fluor}     
\label{findme}                          
\begin{algorithmic} [1]                   
\REQUIRE   $I_{f,\ell}(\mathbf{r})$, $\mathbf{r}_\ell$, $\ell = 1, \ldots, N_{\mathrm{img}}$ 
\STATE initialize $o_f^{(1,0)}(\mathbf{r}) = \sum_\ell I_{f,\ell}(\mathbf{r})/N_{\mathrm{img}}$ 
\STATE initialize $p_f^{(1,0)}(\mathbf{r})$ with all one values
\STATE initialize $\tilde{h}_f(\mathbf{u})$ with the non-aberrated incoherent OTF
\STATE initialize $\mathbf{r}^{(1)}_\ell$ with the scanning position from the registration step
\FOR{$k = 1:K_f$}
	\STATE Sequential gradient descent
	\FOR{$\ell = 1:N_{\mathrm{img}}$}
		\STATE $o_f^{(k,\ell)}(\mathbf{r}) =  o_f^{(k,\ell-1)}(\mathbf{r}) - \nabla_{o_f} f_{f,\ell} (o_f^{(k,\ell-1)}, p_f^{(k,\ell-1)}, \mathbf{r}^{(k)}_\ell) / \max(p_f^{(k,\ell-1)}(\mathbf{r}))^2$
  		\STATE $p_f^{(k,\ell)}(\mathbf{r}) =  p_f^{(k,\ell-1)}(\mathbf{r}) - \nabla_{p_f} f_{f,\ell} (o_f^{(k,\ell-1)}, p_f^{(k,\ell-1)}, \mathbf{r}^{(k)}_\ell) / \max(o_f^{(k,\ell-1)}(\mathbf{r}))^2$
  		\STATE $\xi(\mathbf{u}) = \mathcal{F}\{o_f^{(k,\ell-1)}(\mathbf{r}) \cdot \mathcal{C}\{p_f^{(k,\ell-1)}(\mathbf{r}-\mathbf{r}_\ell)\}\}$
  		\STATE $\tilde{h}_f^{(k,\ell)}(\mathbf{u}) = \tilde{h}_f^{(k,\ell-1)}(\mathbf{u}) - \nabla_{\tilde{h}_f} f_{f,\ell} (o_f^{(k,\ell-1)}, p_f^{(k,\ell-1)}, h_f^{(k,\ell-1)}, \mathbf{r}_\ell^{(k)}) \cdot |\xi(\mathbf{u})|/12[\max(|\xi(\mathbf{u})|) \cdot (|\xi(\mathbf{u})|^2 + \delta)]$, where $\delta$ is chosen to be small
  		\STATE
 	 	\STATE Scanning position refinement
  		\STATE $x^{(k+1)}_\ell$ = $x^{(k)}_\ell - \alpha \nabla_{x_{\ell}} f_{f,\ell} (o_f^{(k,\ell-1)}, p_f^{(k,\ell-1)}, \mathbf{r}^{(k)}_\ell)$
  		\STATE $y^{(k+1)}_\ell$ = $y^{(k)}_\ell - \alpha \nabla_{y_{\ell}} f_{f,\ell} (o_f^{(k,\ell-1)}, p_f^{(k,\ell-1)}, \mathbf{r}^{(k)}_\ell)$
  	\ENDFOR
  	\STATE Nesterov's acceleration
  	\IF{k = 1}
  		\STATE $t_1 = 1$
   		\STATE $o_f^{(k+1,0)}(\mathbf{r}) = o_f^{(k,N_{\mathrm{img}})}(\mathbf{r})$
		\STATE $p_f^{(k+1,0)}(\mathbf{r}) = p_f^{(k,N_{\mathrm{img}})}(\mathbf{r})$
	\ELSE
		\STATE $t_{k+1} = \frac{1+\sqrt{1+ 4t_k^2}}{2}$ 
  		\STATE $o_f^{(k+1,0)}(\mathbf{r}) = o_f^{(k,N_{\mathrm{img}})}(\mathbf{r}) + \frac{t_k -1}{t_{k+1}} \left[o_f^{(k,N_{\mathrm{img}})}(\mathbf{r}) -  o_f^{(k-1,N_{\mathrm{img}})}(\mathbf{r})\right]$
  		\STATE $p_f^{(k+1,0)}(\mathbf{r}) = p_f^{(k,N_{\mathrm{img}})}(\mathbf{r}) + \frac{t_k -1}{t_{k+1}} \left[p_f^{(k,N_{\mathrm{img}})}(\mathbf{r}) -  p_f^{(k-1,N_{\mathrm{img}})}(\mathbf{r})\right]$
  	\ENDIF

\ENDFOR
\end{algorithmic}
\end{algorithm}

\subsection*{B.2. Algorithm for coherent imaging}
For coherent imaging, we initialize $o_c(\mathbf{r})$ with all ones. The pattern, $p_c(\mathbf{r})$, is initialized with the mean of the square root of registered coherent in-focus intensity stack. The pupil function is initialized with a circ function (2D function filled with ones within the defined radius) with the radius defined by the objective NA. In the end, we initialize the scanning position, $\mathbf{r}_{\ell z}$, from the registration of the intensity stacks, $I_{c,\ell z}$, for respective focal planes.

For the coherent imaging reconstruction, we use a total number of $K_c \approx 30$ iterations to converge. We sequentially update $o_c(\mathbf{r})$, $p_c (\mathbf{r})$, $h_c(\mathbf{r})$, and $\mathbf{r}_{\ell}, (\ell = 1, \ldots , N_{\mathrm{img}})$ for each defocused plane (total number of defocused planes is $N_z$) per iteration. Unlike for our fluorescence reconstructions, we do not use the extra Nesterov's acceleration step in the QP reconstruction.

\begin{algorithm} [H]             
\caption{Coherent imaging reconstruction}
\label{algo_coh}     
\label{findme}                          
\begin{algorithmic} [1]                   
\REQUIRE   $I_{c,\ell z}(\mathbf{r})$, $\mathbf{r}_{\ell z}$, $\ell = 1, \ldots, N_{\mathrm{img}}$ 
\STATE initialize $o_c^{(1,0)}(\mathbf{r})$ with all one values 
\STATE initialize $p_c^{(1,0)}(\mathbf{r}) = \sum_{\ell} \sqrt{I_{c,\ell, z=0}(\mathbf{r}+\mathbf{r}_{\ell, z=0})}/N_{\mathrm{img}}$ 
\STATE initialize $\tilde{h}_c^{(1,0)}(\mathbf{u})$ with all one values within a defined radius set by the objective NA
\STATE initialize $\mathbf{r}^{(1)}_{\ell z}$ with the scanning position from the registration step
\FOR{$k = 1:K_c$}
	\STATE Sequential gradient descent
	\FOR{$t = 1:(N_{\mathrm{img}} \cdot N_z) $}
		\STATE $z = z_{\mathrm{mod}(t,2)}$
		\STATE $\ell = \mathrm{mod}(t,N_{\mathrm{img}})$
		\IF {$t < N_{\mathrm{img}} \cdot N_z$}
			\STATE $o_{c}^{(k,t)}(\mathbf{r}) =  o_{c}^{(k,t-1)}(\mathbf{r}) - \nabla_{o_c} f_{c,\ell z} (o_c^{(k,t-1)}, p_c^{(k,t-1)}, \tilde{h}_c^{(k,t-1)}, \mathbf{r}^{(k)}_{\ell z}) / \max\left(\left|p_c^{(k,t-1)}(\mathbf{r})\right|\right)^2$
  			\STATE $p_{c}^{(k,t)}(\mathbf{r}) =  p_{c}^{(k,t-1)}(\mathbf{r}) - \nabla_{p_c} f_{c,\ell z} (o_c^{(k,t-1)}, p_c^{(k,t-1)}, \tilde{h}_c^{(k,t-1)}, \mathbf{r}^{(k)}_{\ell z}) / \max\left(\left|o_c^{(k,t-1)}(\mathbf{r})\right|\right)^2$
  			\STATE $\xi(\mathbf{u}) = \mathcal{F}\{o_c^{(k,t-1)}(\mathbf{r})\cdot \mathcal{C}\{p_c^{(k,t-1)}(\mathbf{r}-\mathbf{r}_\ell)\}\}$
  			\STATE $\tilde{h}_c^{(k,t)}(\mathbf{u}) = \tilde{h}_c^{(k,t-1)}(\mathbf{u}) - \nabla_{\tilde{h}_c} f_{c,\ell z} (o_c^{(k,t-1)}, p_c^{(k,t-1)}, \tilde{h}_c^{(k,t-1)}, \mathbf{r}^{(k)}_{\ell z}) \cdot |\xi(\mathbf{u})|/5[\max(|\xi(\mathbf{u})|) \cdot (|\xi(\mathbf{u})|^2 + \delta)]$, where $\delta$ is chosen to be small 
  		\ELSE
  			\STATE Do the same update but save to $o_{c}^{(k+1,0)}(\mathbf{r}), p_{c}^{(k+1,0)}(\mathbf{r}), \tilde{h}_{c}^{(k+1,0)}(\mathbf{r})$
  		\ENDIF 
  		\STATE
 	 	\STATE Scanning position refinement
  		\STATE $x^{(k+1)}_{\ell z}$ = $x^{(k)}_{\ell z} - \beta \nabla_{x_{\ell z}} f_{c,\ell z} (o_c^{(k,t-1)}, p_c^{(k,t-1)}, \tilde{h}_c^{(k,t-1)}, \mathbf{r}^{(k)}_{\ell z})$
  		\STATE $y^{(k+1)}_{\ell z}$ = $y^{(k)}_{\ell z} - \beta \nabla_{y_{\ell z}} f_{c,\ell z} (o_c^{(k,t-1)}, p_c^{(k,t-1)}, \tilde{h}_c^{(k,t-1)}, \mathbf{r}^{(k)}_{\ell z})$
  	\ENDFOR
\ENDFOR
\end{algorithmic}
\end{algorithm}

\section*{Appendix C: Sample preparation}

Results presented in this work targeted super-resolution of 1 $\mu$m and 2 $\mu$m diameter polystyrene microspheres (Thermofischer) that were fluorescently tagged to emit at a center wavelength of $\lambda_{\mathrm{em}}=605$ nm. Monolayer samples of these microspheres were prepared by placing microsphere dilutions (60 uL stock-solution/500 uL isopropyl alcohol)  onto \#1.5 coverslips and then allowing to air-dry. High-index oil ($n_m(\lambda)=1.52$ at $\lambda=532$ nm) was subsequently placed on the coverslip to index-match the microspheres. An adhesive spacer followed by another \#1.5 coverslip was placed on top of the original coverslip to assure a uniform sample layer for imaging.

\section*{Appendix D: Posedness of the problem}
\label{sec_AppD}

In this paper, we illuminate the sample with an unknown speckle pattern to encode both large-FOV and high-resolution information into our measurement. To decode the high-resolution information, we need to jointly estimate the speckle pattern and the sample. This framework shares similar characteristics with the work on blind SIM first introduced by~\cite{Mudry2012}, where completely random speckle patterns were sequentially illuminated onto the sample. Unfortunately, the reconstruction formulation proposed in that work is especially ill-posed due to randomness between the illumination patterns, \textit{i.e.}, if $N_{\mathrm{img}}$ raw images are taken, there would be $N_{\mathrm{img}}+1$ unknown variables to solve for ($N_{\mathrm{img}}$ illumination patterns and 1 sample distribution). To better condition this problem, priors based on speckle statistics~\cite{Mudry2012, Ayuk2013,Jost2015, Negash2016,Yeh2017} and sample sparsity~\cite{Min2013, Labouesse2016} can be introduced, pushing blind SIM to 2$\times$ resolution gain. However, to implement high-content microscopy using SIM, we desire a resolution gain of $>2\times$. Even with priors, we found that this degree of resolution gain was not experimentally achievable with uncorrelated and random speckle illuminations, due to the reconstruction formulation being so ill-posed. 

In this work, we improve the posedness of the problem by illuminating with a translating speckle pattern, as opposed to randomly changing speckle patterns. Because each individual illumination pattern at the sample is a laterally shifted version of every other illumination pattern, the posedness of the reconstruction framework dramatically increases. Previous works~\cite{Dong2014, Yilmaz2015, kaikai2018} have also demonstrated this concept to effectively achieve beyond $2\times$ resolution gain. 

\section*{Appendix E: Self-calibration analysis}

In Sec.~\ref{subsec_inverseproblem} and~\ref{subsec_inverseproblem_ph}, we presented the inverse problem formulation for super-resolution fluorescence and QP. We note that those formulations also included terms to self-calibrate for unknowns in the system's experimental OTF and the illumination pattern's scan-position. Here we demonstrate how these calibrations are important for our reconstruction quality.

\begin{figure}[tbh]
\centering
\includegraphics[width=13cm]{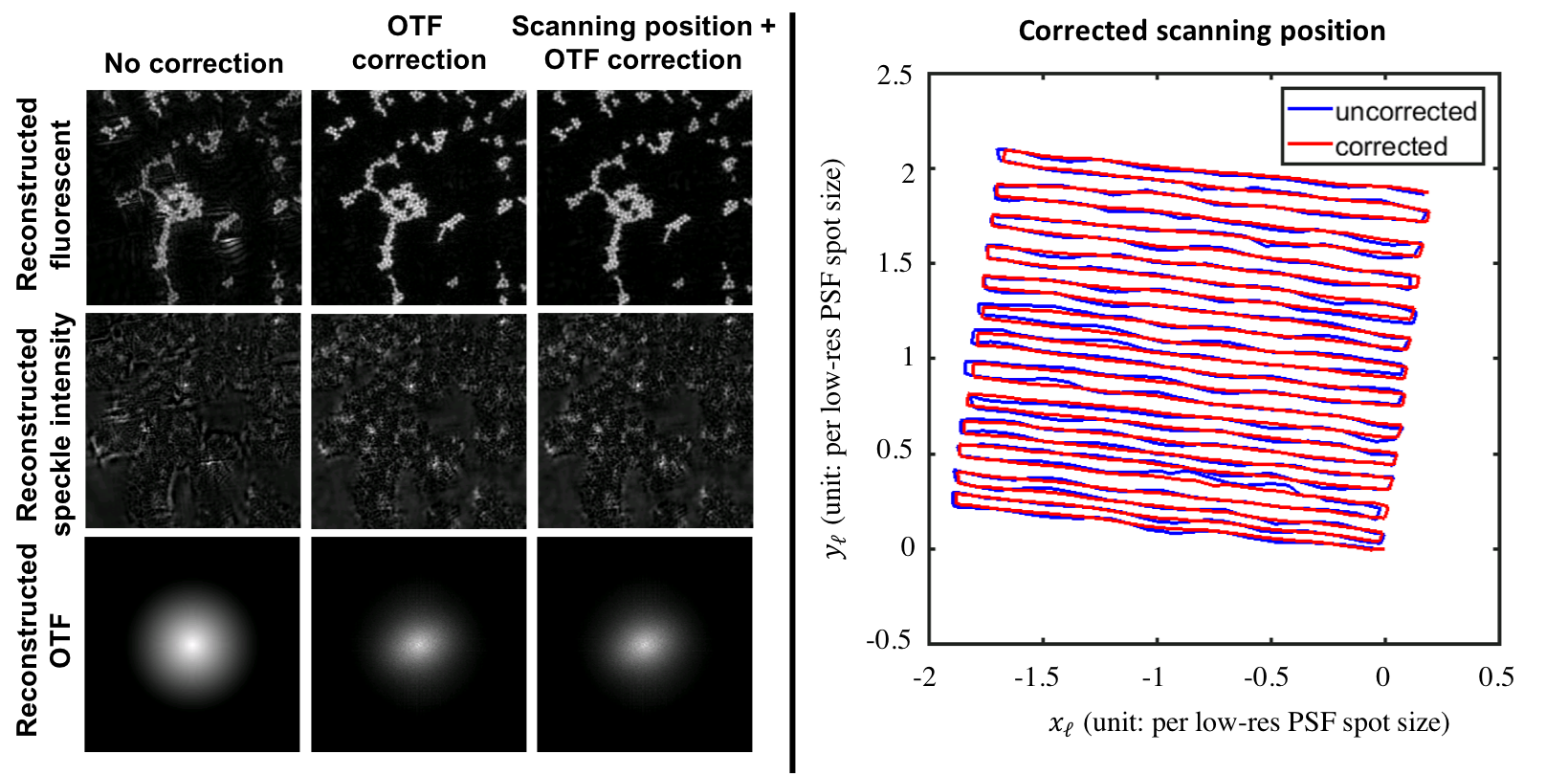}
\caption{Algorithmic self-calibration significantly improves fluorescence super-resolution reconstructions. Here, we compare the resconstructed fluorescence image, speckle intensity, and OTF with no correction, OTF correction, and both OTF correction and scanning position correction. The right panel shows the overlay of the uncorrected and corrected scanning position trajectories.}
\label{fig_SC_FL}
\end{figure} 

To demonstrate the improvement in our fluorescence imaging reconstruction due to the self-calibration algorithm, we select a region of interest from the dataset presented in Fig.~\ref{fig_large_FOV_FL}. Figure~\ref{fig_SC_FL} shows the comparison of the SR reconstruction with and without self-calibration. The SR reconstruction with no self-calibration contains severe artifacts in reconstructions of both the speckle illumination pattern and the sample's fluorescent distribution. With OTF correction, dramatic improvements in the fluorescence SR image are evident. OTF correction is  especially important when imaging across a large FOV (Fig.~\ref{fig_large_FOV_FL} and~\ref{fig_large_FOV_FL_QP}) due to space-varying aberrations. 

Further self-calibration to correct for errors in the initial estimate of the illumination pattern's trajectory enables further refinement of the SR reconstruction. We see that this illumination trajectory demonstrates greater smoothness after undergoing self-calibration. We fully expect that this calibration step to have important ramifications in cases where the physical translation stage is of lower stability or more inaccurate incremental translation. 

\begin{figure}[tbh]
\centering
\includegraphics[width=13cm]{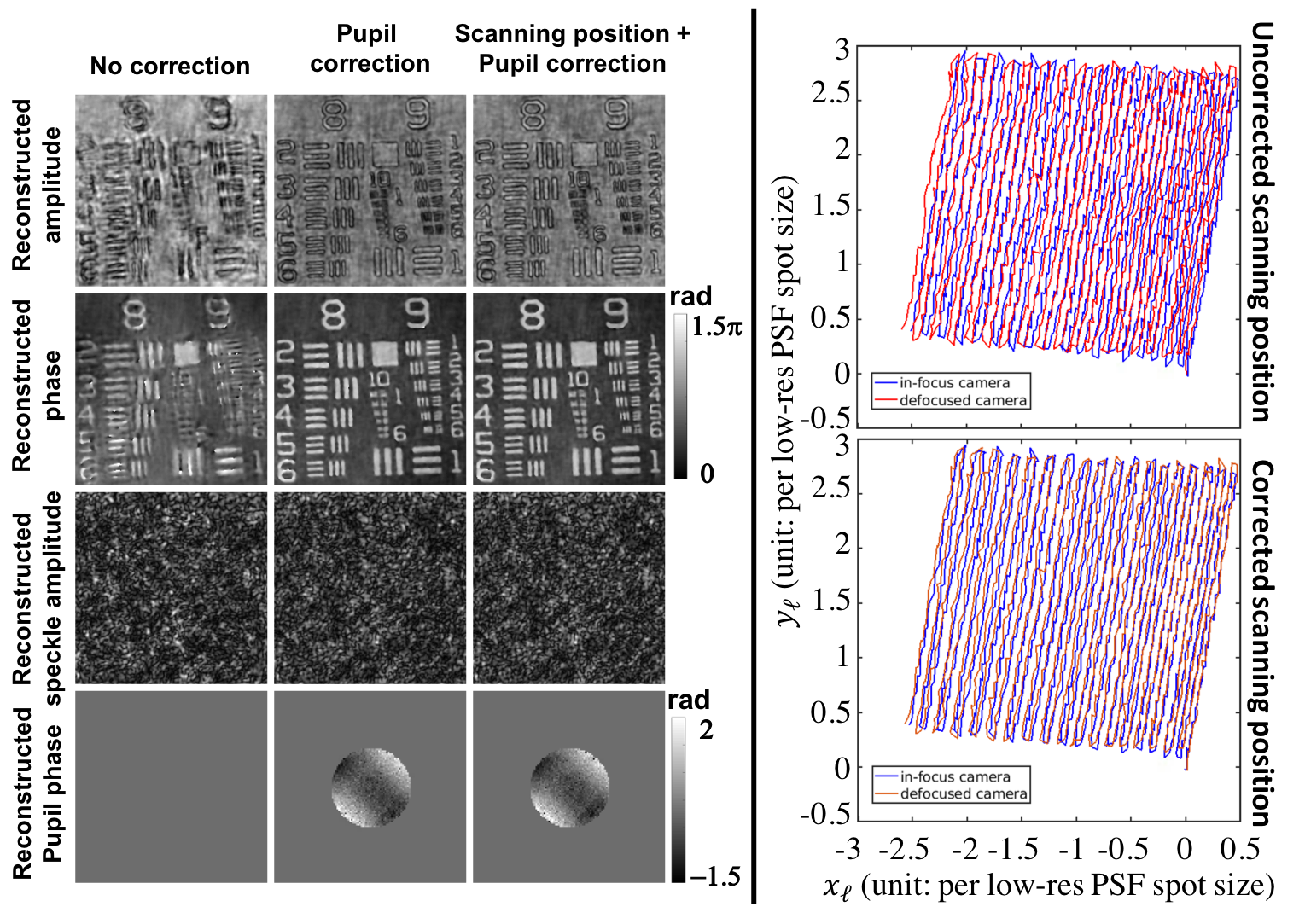}
\caption{Algorithmic self-calibration significantly improves coherent super-resolution reconstructions. We show a comparison of reconstructed amplitude, phase, speckle amplitude, and phase of the pupil function with no correction, pupil correction, and both pupil correction and scanning position correction. The right panel shows the overlay of scannning position trajectory for the in-focus and defocused cameras before and after correction.}
\label{fig_SC_PH}
\end{figure} 

We also test how the self-calibration affects our phase reconstruction, using the same dataset as in Fig.~\ref{fig_verify_phase}. Similar to the conclusion from the fluorescence self-calibration demonstration, pupil correction (coherent OTF) plays an important role in reducing SR reconstruction artifacts as shown in Fig.~\ref{fig_SC_PH}. The reconstructed pupil phase suggests that our system aberration is mainly caused by astigmatism. Further refinement of the trajectory of the illumination pattern improves the SR resolution by resolving one more element (group 9 element 6) of the USAF chart. Paying more attention to the uncorrected and corrected illumination trajectory, we find that the self-calibrated trajectory of the illumination pattern tends to align the trajectories from the two coherent cameras. We also notice that the trajectory from the quantitative-phase channels seems to jitter more compared to the fluorescence channel. We hypothesize that this is due to longer exposure time for each fluorescence acquisition, which would average out the jitter. 

\section*{Funding}

Gordon and Betty Moore Foundation's Data-Driven Discovery Initiative (GBMF4562); Ruth L. Kirschstein National Research Service Award (F32GM129966).

\end{document}